\documentclass[fleqn,usenatbib]{mnras}

\usepackage{newtxtext,newtxmath}

\usepackage[T1]{fontenc}

\DeclareRobustCommand{\VAN}[3]{#2}
\let\VANthebibliography\thebibliography
\def\thebibliography{\DeclareRobustCommand{\VAN}[3]{##3}\VANthebibliography}

\usepackage{graphicx}	
\usepackage{amsmath}	
\usepackage{lscape}

\usepackage[dvipsnames]{xcolor} 

\usepackage{xspace}

\newcommand{\be}{\begin{equation}}
\newcommand{\ee}{\end{equation}}
\newcommand{\bea}{\begin{eqnarray}}

\newcommand{\eea}{\end{eqnarray}}
\DeclareMathAlphabet\mathbfcal{OMS}{cmsy}{b}{n}

\title[Structure of the planetary 2:1 MMR.]{Mapping the structure of the planetary 2:1 mean motion resonance. \\ The TOI-216, K2-24, and HD27894 systems.}

\author[C. A. Giuppone et al.]{{Cristian} {Giuppone}$^{1}$ \thanks{cristian.giuppone@unc.edu.ar}
, {{Adrián} {Rodríguez}$^{2}$}
, {{Viviam} {Alencastro}$^{3}$}
, *{{Fernando} {Roig}$^{3}$}\thanks{froig@on.br}
, {{Tabaré} {Gallardo}$^{4}$}
\\
$^{1}${{Observatorio Astrónomico - IATE}, {Universidad Nacional de Córdoba}, {{Laprida 854}, {Córdoba}, {X5000BGR}, {Córdoba}, {Argentina}}} \\
$^{2}${{Observatório do Valongo}, {Universidade Federal do Rio de Janeiro}, {{Ladeira Pedro Antônio 43}, {Rio de Janeiro}, {20080-090}, {RJ}, {Brazil}}} \\
$^{3}${{Observatório Nacional}, {Ministério da Ciência, Tecnologia e Inovaçoes}, {{Rua Gen José Cristino 77}, {Rio de Janeiro}, {20921-400}, {RJ}, {Brazil}}} \\
$^{4}${{Facultad de Ciencias}, {Universidad de la República}, {{Iguá 4225}, {Montevideo}, {11400}, {Uruguay}}} \\
}

\date{Accepted 2023. Received 2022; in original form 2022}

\pubyear{2022}

\begin{document}
\label{firstpage}
\pagerange{\pageref{firstpage}--\pageref{lastpage}}
\maketitle

\begin{abstract}
{Mean motion resonances (MMR) are a frequent phenomenon among extrasolar planetary systems. Current observations indicate that many systems have planets that are close to or inside the 2:1 MMR, when the orbital period of one of the planets is twice the other. Analytical models to describe this particular MMR can only be reduced to integrable approximations in a few specific cases. While there are successful approaches to the study of this MMR in the case of very elliptic and/or very inclined orbits using semi-analytical or semi-numerical methods, these may not be enough to completely understand the resonant dynamics.
In this work, we propose to apply a well-established numerical method to assess the global portrait of the resonant dynamics, which consists in constructing dynamical maps. Combining these maps with the results from a semi-analytical method, helps to better understand the underlying dynamics of the 2:1 MMR, and to identify the behaviors that can be expected in different regions of the phase space and for different values of the model parameters.
We verify that the family of stable resonant equilibria bifurcate from symmetric to asymmetric librations, depending on the mass ratio and eccentricities of the resonant planets pair. This introduces new structures in the phase space, that turns the classical V-shape of the MMR, in the semi-major axis vs. eccentricity space, into a sand clock shape. We construct dynamical maps for three extrasolar planetary systems, TOI-216, HD27894, and K2-24, and discuss their phase space structure and their stability in the light of the orbital fits available in the literature.}
\end{abstract}

\begin{keywords}
{numerical methods, extrasolar planets, dynamical evolution and stability, mean motion resonance}
\end{keywords}

\section{Introduction}\label{sec:intro}

Mean motion resonances (MMR) are a frequent phenomenon among extrasolar planetary systems. The occurrence of MMR provides important clues about the formation and early evolution of such systems. In particular, capture or closeness of planets pairs to MMR is usually taken as an evidence of some sort of primordial radial migration of the planets. 

Current observations indicate that many systems have planets that are close to or inside the 2:1 MMR, where the orbital period of one of the planets is twice the other. Analyzing the distribution of planets pairs near the 2:1 MMR that have mass estimates, taken from the Extrasolar Planets Encyclopaedia (\url{http://exoplanet.eu/}), we can verify that there is a higher frequency of pairs with a more massive outer planet. Systems with mass ratios (outer/inner) greater than 2 are more prevalent for periods ratios (outer/inner) greater than 2, while systems with mass ratios smaller than 0.5 occur more often for period ratios smaller than 2.  Also, systems with mass ratios close to 1 are more often detected. Biases in the different detection methods \citep[e.g.][]{Giuppone2009} may indicate that the actual number of resonant systems is significantly larger than currently observed.

The distribution of resonant pairs also shows a gap in period ratios at the exact 2:1 MMR, with an overpopulation of planets with period ratios between 2.0 and 2.05. \citet{Goldreich+2014} proposed that this shift towards slightly larger period ratios reflects the same asymmetry that requires convergent migration for resonance capture. Possible evidence of subsequent tidal evolution to explain the dispersion of period ratios around the 2:1 MMR has been addressed by \citet{Delisle2014,Delisle2014A} and \citet{Ramos+2017}. Other mechanisms to explain this dispersion rely on interaction with disks of planetesimals \citep{Chatterjee2015}, or evolution in turbulent disks \citep{Batygin2017}.

The radial velocity (RV) and transit timing variations (TTV) techniques are very successful in characterizing the orbits and masses of extrasolar planets in MMR from observational data \citep[][]{Bean2009, Rivera2010, Trifonov2014}. However, planet characterization applying these techniques to sparse data, or to data with incomplete phase coverage, sometimes led to ill-determined system architectures \citep[e.g.][]{Beauge2008, dawson19, kipping19, dawson21}. In the case of the 2:1 MMR, for example, the combined RV signal of a pair of planets in low eccentricity orbits could be misinterpreted as a single planet with moderate eccentricity if the data are sparse \citep{Escude2010, Boisvert+2018}, or even with a coorbital pair in a trojan configuration \citep{Giuppone2012}. 

Analytical models to describe the 2:1 MMR have the limitation of the many degrees of freedom of the system, which can only be reduced to integrable approximations in a few specific cases. In particular, analytical models describing MMR between two planets have been developed for coplanar orbits \citep{Beauge2003, Batygin2013b, Deck2013}. There have been some successful approaches using semi-analytical or semi-numerical methods, yet these are not enough to completely understand the resonant dynamics. For the 2:1 MMR, some models provide families of apsidal corrotation resonances \citep[ACR;][]{Beauge2006, Michtchenko2008a, Michtchenko2008b, Voyatzis2009}. A recent work introduces a semi-analytical model that allows to study MMR for any eccentricities and inclinations of the planets \citep{Gallardo2021}.

In this work, we propose to apply a well-established numerical method to assess the global portrait of the 2:1 MMR dynamics, which consists in constructing dynamical maps. Combining these maps with semi-analytical methods may help to better understand the underlying resonant structure, allowing to identify the different behaviors that can be expected in different regions of the phase space, and for different values of the model parameters.

We aim to study some specific systems near the 2:1 MMR ---period ratios of $2 \pm 0.05$, based on the offset produced by tidal evolution of the planets after disk dispersal \citep{Ramos+2017}---, and to discuss their dynamics in the neighborhood of their best fit orbital parameters. 
We wish to detect the regions that may or may not harbor planets and apply this to constrain the orbital parameters that are usually unconstrained from the observations, like the pericenters. We also aim to explore other configurations beyond the orbital fits in order to understand how and why the resonance portrait depends on the parameters of the system. This, in turn, may help us to understand why the real system is in its current configuration.
In particular, we choose three systems with masses of the planets pairs that cover a wide range of mass ratios: TOI-216 $(m_2/m_1=9.49)$, HD-27894 $(m_2/m_1=0.24)$, and K2-24 $(m_2/m_1=0.81)$. Along this work, we will use the sub-indexes 1 and 2 to refer to the inner and outer planet, respectively. 

The paper is organized as follows: in Section \ref{methods} we explain the methodology that will be used, providing details on the construction of the dynamical maps and the application of the semi-analytical model of \citet{Gallardo2021}, hereafter GBG21. In Section \ref{systems}, we present the analysis of the three extrasolar systems referred above (Sect. \ref{sec:TOI-216}, \ref{sec:HD27894} and \ref{sec:K2-24}), as well as some general results concerning the structure of the 2:1 MMR (Sect. \ref{sec:sand-clock}). Finally, our conclusions are summarized in Section \ref{sec:conclusions}.

\section{Methodology}\label{methods}

The 2:1 MMR between a pair of planets is characterized by the two resonant or critical angles
\begin{equation}
    \sigma_1 =\lambda_1 -2\lambda_2 +\varpi_1, \qquad
    \sigma_2 =\lambda_1 -2\lambda_2 +\varpi_2,
\end{equation}
where $\lambda_i$ are the mean longitudes of the planets and $\varpi_i$ are the longitudes of periastron. 
We assume that at least one of these critical angles must librate around a stable equilibrium value in order to have the system in the 2:1 MMR. The libration of $\sigma_i$ is coupled with an oscillation of the semi-major axis $a_i$ around the resonant value $a_\mathrm{res}$. The resonant domain is characterized by the maximum libration amplitude $\delta\sigma_i$, or equivalently, $\delta a_i$, which defines the so-called separatrix of the MMR. The secular angle
\begin{equation}
    \Delta\varpi =\varpi_2 -\varpi_1 = \sigma_2 -\sigma_1
\end{equation}
is also a relevant parameter to characterize the dynamics of the MMR.
In particular, when both critical angles are librating at the same time, then $\Delta\varpi$ will also librate, leading to a configuration usually referred to as apsidal corotation resonance or ACR.

In this work, we employ the technique of dynamical maps in order to investigate the resonant structure of the 2:1 MMR. 
Usually, a dynamical map is constructed as a 2D grid of numerical simulations by varying the initial value of two orbital elements of a selected body, keeping the remaining parameters unchanged. The most usual representation of a dynamical map to study the structure of a mean-motion resonance is based on a grid of initial semi-major axis and eccentricity or inclination. 

We construct dynamical maps considering grids in the semi-major axis vs. eccentricity plane, around the nominal position of the less massive planet in a system with only two planets close to the 2:1 MMR. We choose to vary the parameters of the less massive planet over the grids under the assumption that its orbit is more prone to higher variations due to the perturbation of the more massive companion. 
Each grid includes $150\times150$ numerical simulations, using two different initial configurations of the periastron longitudes, $\Delta\varpi=0^{\circ}$ and $180^{\circ}$; thus we present the maps in the $a_i, e_i \cos\Delta\varpi$ plane. The initial semi-major axis and eccentricity of the more massive planet are taken to be equal to the nominal values of the best fit for each system analyzed (Sect. \ref{systems}). The initial mean anomalies, inclinations and ascending nodes are assumed to be zero. 

The simulations are carried out in the framework of the general planar three body problem. We numerically solved the full equations of motion in Newtonian form, together with the corresponding variational equations of the system. We use a version of the Bulirsh Stoer integrator \citep{NumRecip1992}, with adaptive step size, that has been modified to independently monitor the error in each variable. We impose a relative precision better than $10^{-13}$. A simulation is stopped when the distance between the star and any planet is less than their mutual radii, or when a planet is ejected from the system (astrocentric distance $>2$~au) after scattering with the other bodies.
We integrate each numerical simulation over a time span of 100 or 200 yr, depending on the specific system, which represents between $1\,000$ and $2\,500$ orbital periods of the outer planet. 

For each numerical simulation on the grid, we compute several dynamical indicators allowing us to identify the regions of stable and unstable motion, or regular and chaotic motion, as well as some properties of the system such as the libration center and the separatrix of the resonance. The indicators we consider are: the maximum variation of the semi-major axis, $\max\Delta a_i$, the maximum variation of eccentricity, $\max\Delta e_i$, and the maximum amplitude of the critical angle, $\max\Delta\sigma_i$\footnote{The notation $\Delta\sigma_i$ refers to the libration amplitude of a specific orbit, while $\delta\sigma_i$ denotes the resonance width. Therefore, $\max\Delta\sigma_i\approx\delta\sigma_i$.}. All these dynamical indicators are coded in a suitable color scale.
In Sects. \ref{sec:TOI-216} and \ref{sec:HD27894}, we present and discuss the maps of the dynamical indicators corresponding to the less massive planet in the system. Similar maps with the dynamical indicators corresponding to the more massive planet are presented for completeness in the Appendix. In Sect. \ref{sec:K2-24}, on the other hand, we present and discuss the indicators of both planets, because they have similar masses.

We also compute for each simulation the MEGNO chaos indicator, $\langle Y^*\rangle =\log_{10}\left(\lvert\langle Y\rangle -2\rvert \right)$ \citep[see][for details]{Cincotta.Simo.1999}.
Since the value of this indicator may be sensitive to the integration time span \citep[][]{Cincotta2003, Hinse2010}, we took the TOI-216 system as a test case and computed the maps of MEGNO for time intervals of 100, 1\,000 and 10\,000 years. The results are presented in the Appendix. {We verified that the maps do not display significant differences, although the final value of the MEGNO may actually differ in certain regions. In principle, we may assume that the time span of 100-200 years we used in all the simulations provides a MEGNO value that allows the identification of regions of strong chaos, but for the regions of regular motion, the results should be treated with caution.}

In addition to the dynamical maps, we also use the model of GBG21, which is based on a semi-analytical approach for planetary resonances around single or binary stars. In the framework of Jacobi and Poincaré variables, this model computes the numerical value of the averaged resonant disturbing function, allowing to predict the strength of any MMR, the location and stability of the equilibrium points, the libration period of the resonant angle, and the width of the resonant domain. We apply the model of GBG21 to estimate the position of equilibrium points and the width of the 2:1 MMR, as functions of the orbital eccentricity. These results are then compared to the results from the dynamical maps, allowing us to better understand the global behavior of the resonant dynamics of the planetary systems under analysis. 

\begin{figure}
    \includegraphics[width=\columnwidth, trim=0 0 0 2.3cm, clip]{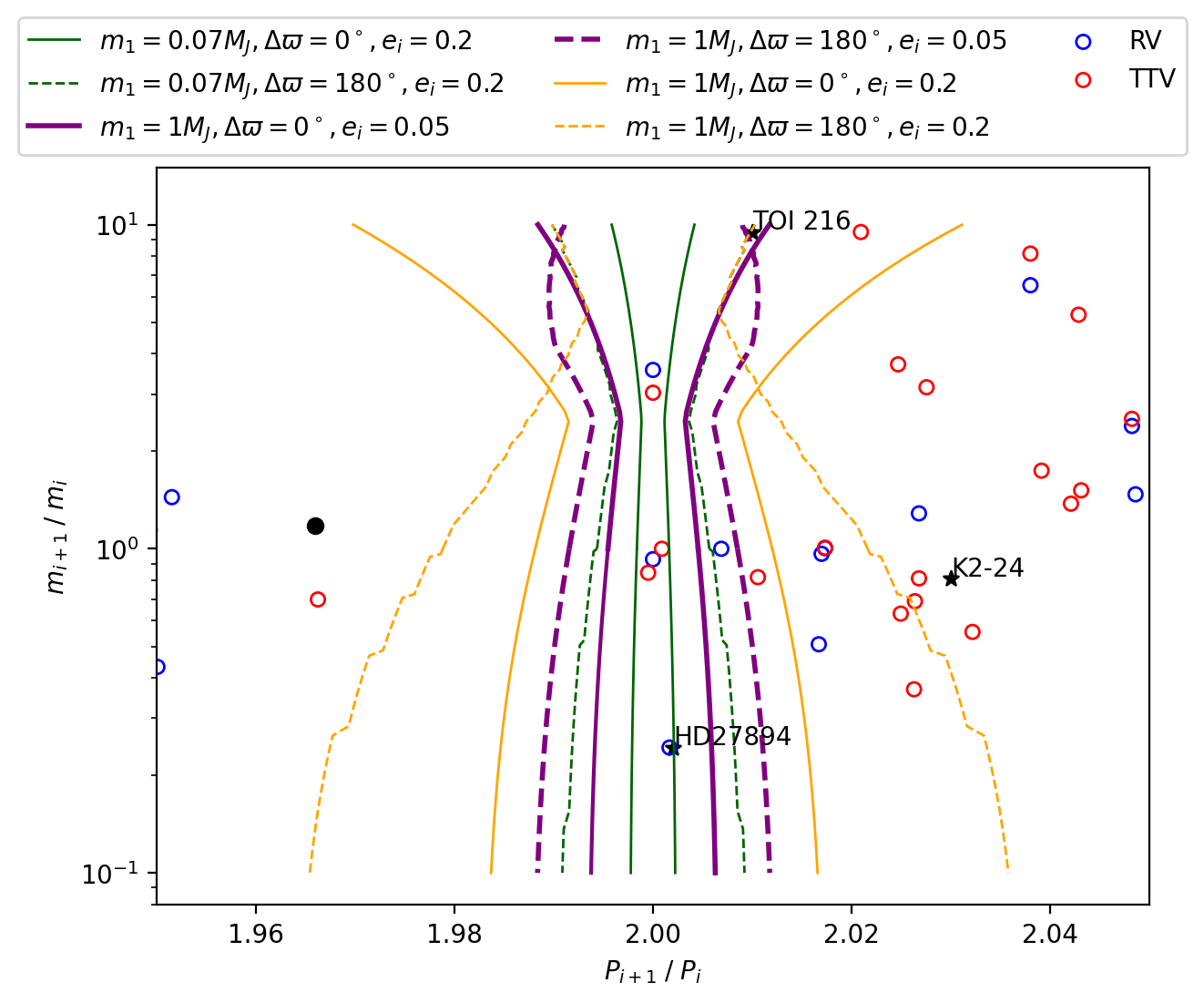}
    \caption{The 2:1 MMR width, $\delta a$, represented in the space of mass ratio vs. period ratio, computed using the semi-analytic model of GBG21. Full and dashed lines correspond to $\Delta\varpi=0^\circ$ or $180^\circ$, respectively. Each line color corresponds approximately to each of the three systems analyzed in this work: TOI-216 (green) HD27894 (magenta), and K2-24 (orange). These systems are indicated by black stars. Other planets pairs in the 2:1 MMR are indicated by open circles depending on the discovery technique: radial velocity (blue) or transit timing variations (red). The Uranus-Neptune pair is indicated by a full black dot.}
    \label{fig:names}
\end{figure}

Figure \ref{fig:names} shows the maximum widths of the 2:1 MMR, in the representative plane of period ratio, $P_2/P_1$, vs. mass ratio, $m_2/m_1$, for a pair of adjacent planets. The width corresponds to the maximum stable region of the resonant libration in the semi-major axis domain, $\delta a$ (or equivalently $\delta P$), for the less massive planet in the system. To apply the model of GBG21, we fix the mass $m_1$ and the semi-major axis and of the innermost planet, $a_1=0.15$ au, and the orbital eccentricities which equal for both planets. Then, we calculate $\delta a_i$ by varying the mass $m_2$ and semi-major axis $a_2$ of the outermost planet over the adopted ranges of $a_2/a_1$ (or equivalently $P_2/P_1$) and $m_2/m_1$. The planets are assumed to be coplanar and the longitudes of pericenter are also fixed in such a way that $\Delta\varpi=0^{\circ}$ (solid lines), or $\Delta\varpi=180^{\circ}$ (dashed lines). 

Figure \ref{fig:names} also shows the distribution of several known extrasolar planetary systems. Open circles stand for planets pairs in the vicinity of the 2:1 MMR characterized through radial velocity (RV, in blue) and through transit timing variations (TTV, in red). 
The full black dot is the Uranus-Neptune system.
The star symbols represent the three systems that we investigate in this work, namely, TOI-216, K2-24 and HD27894. It is worth noting that the position of the open circles in the plot should not be directly compared to the widths of the 2:1 MMR, since $\delta a_i$ was obtained for fixed values of $m_1$, $a_1$, and $e_1=e_2$ that can significantly differ from the values of the real systems. 

On the other hand, the star symbols can be better compared to the widths, because $\delta a_i$ have been computed for values similar to those of TOI-216 (green lines), HD27894 (magenta lines), and K2-24 (orange lines). The comparison indicates that TOI-216 would be in the 2:1 MMR provided that $\Delta\varpi=180^{\circ}$, but this conclusion may be biased by the fact that $e_1$ differs significantly from $e_2$ in this system. On the other hand, HD27894 is expected to be inside the 2:1 MMR, while K2-24 would not be inside the 2:1 MMR.

Taking into account the fixed values of $m_1$ and $e_1$ that we have used to compute the widths in Fig. \ref{fig:names} for the three systems, we can conclude that $\delta a_i$ increases with $m_1$ and also with $e_1$. In addition, $\delta a_i$ is higher when $\Delta\varpi=180^{\circ}$ than when $\Delta\varpi=0^{\circ}$. On the other hand, $\delta a_i$ also varies with the mass ratio, and displays a minimum at $m_2/m_1\approx 2.5$.

\begin{figure*}
	\includegraphics[width=0.3\textwidth,angle=270]{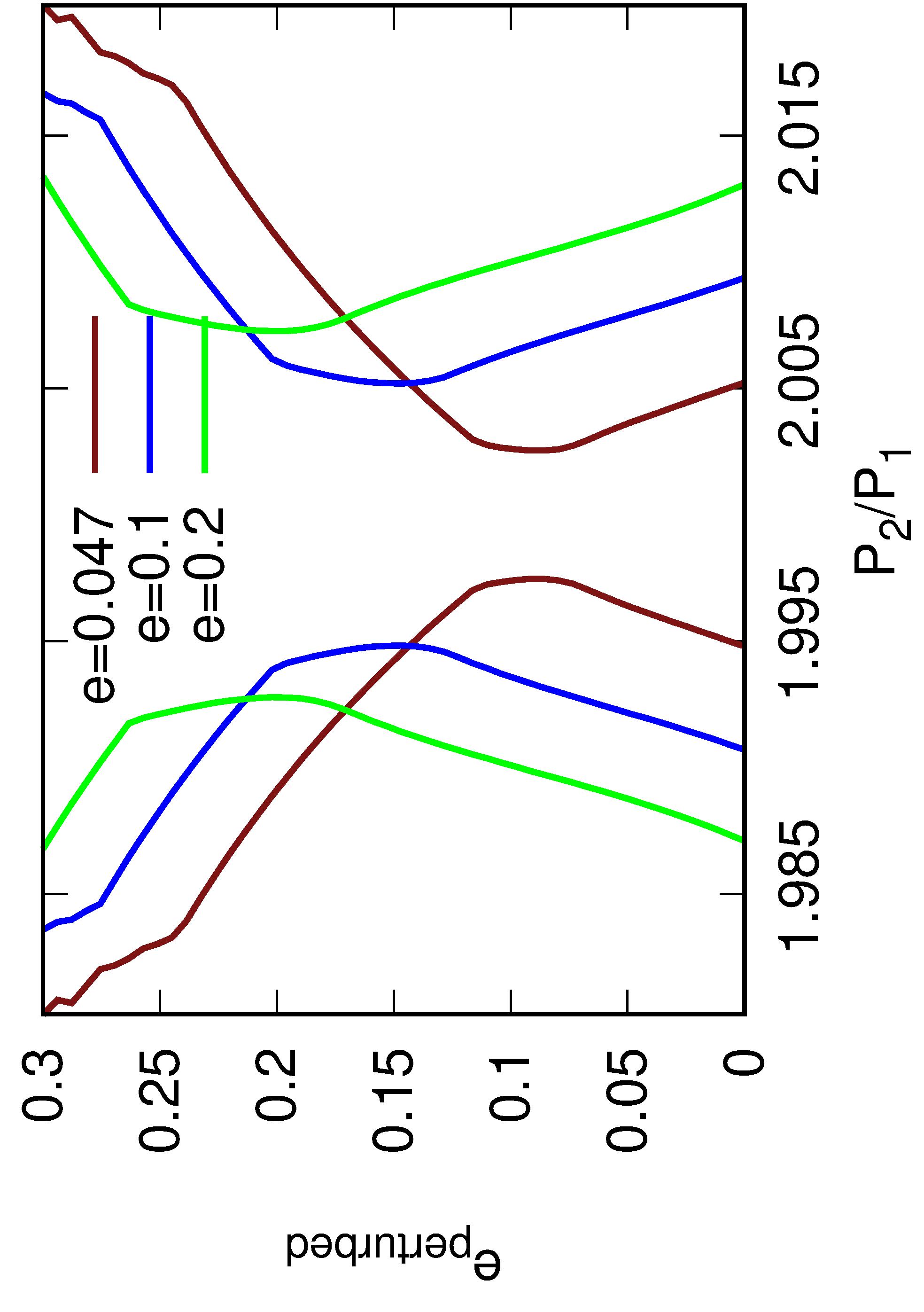}
	\includegraphics[height=0.45\textwidth,angle=270]{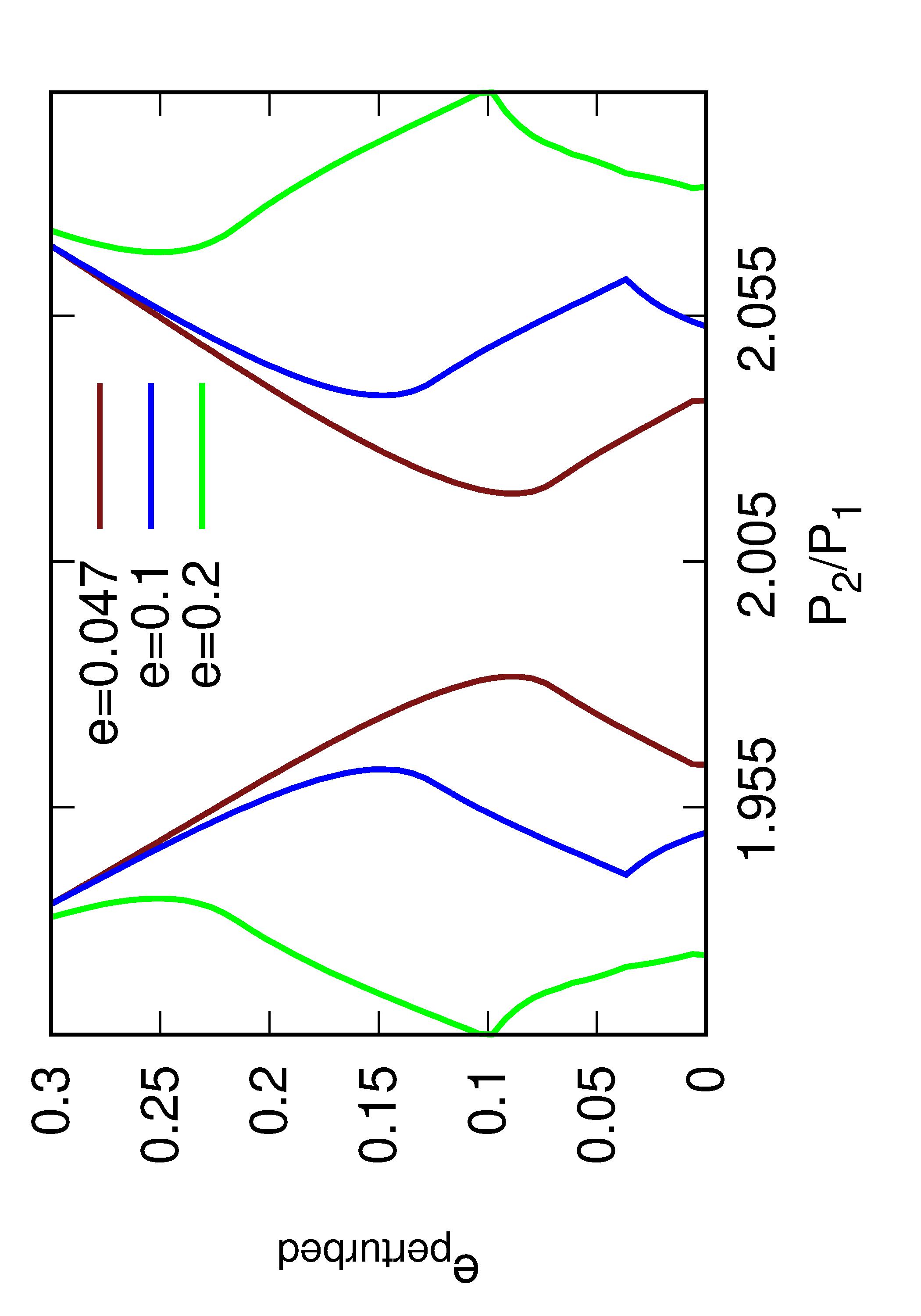}
    \caption{The 2:1 MMR width, $\delta a$, represented in the space of period ratio vs. eccentricity of the less massive (perturbed) planet, $m_2$, computed using the semi-analytic model of GBG21. The results correspond to the HD27894 system with  $\Delta\varpi=0^\circ$. Each color line corresponds to different values of the eccentricity of the more massive (perturbing) planet, $m_1$. Left: assuming the actual mass of $m_1$. Right: assuming a mass $m_1$ ten times larger. See text for details.}
    \label{fig:deltaa}
\end{figure*}

The semi-analytical model also allows to predict the variation of $\delta a_i$ as a function of the eccentricities $e_1,e_2$. This is exemplified in Fig. \ref{fig:deltaa} for the HD27894 system. The left panel shows the resonance width $\delta a_2$ in the $e_2,P_2/P_1$ plane, that corresponds to the HD27894c planet (the perturbed one), assuming three different values for the eccentricity of the HD27894b planet (the perturbing one), $e_1=0.047,\,0.1,\,0.2$. At variance with the classical V-shape that is characteristic of the resonances in the restricted circular three body problem, here we observe something that resembles a sand clock shape. The overall width increases with $e_1$, while the neck of the sand clock
(the narrowest part)
shifts to higher values of $e_2$. The overall width
(for the same $e_1$)
also increases with the mass $m_1$ of the perturbing planet, as we can see in the right panel of Fig. \ref{fig:deltaa}. A more in depth discussion about this sand clock shape will be provided along the following sections.

\section{Results}\label{systems}

\subsection{TOI-216}\label{sec:TOI-216}

TOI-216 is an 11.5 TESS apparent magnitude, main-sequence K-dwarf with $T_\text{eff} = 5045\pm110$~K and $\mathrm{[Fe/H]}=-0.16 \pm 0.09$~dex \citep{dawson19}. The system hosts a pair of warm, large exoplanets discovered by the TESS mission near the mutual 2:1 MMR, and both of them exhibit TTVs. \citet{kipping19} and \citet{dawson19} presented two independent orbital fits that cannot constrain the planetary masses and eccentricities directly from the TTV data. Moreover, the authors provide different values of the stellar parameters that affect the properties of the planets. We choose to study this system configuration because it represents a good example of a large $m_2/m_1$ mass ratio system.

\begin{table}
	\centering
	\caption{Dynamical parameters of the TOI-216 system, according to different fits: \citet{dawson19} - D2019, \citet{kipping19} - K2019, and  \citet{dawson21} - D2021.}
	\label{tab:toi216}
    \renewcommand{\arraystretch}{1.4}
	\begin{tabular}{lrrr}
        \hline
        Data & {D2019} & K2019 & D2021 \\
        \hline
        $m_1$ ($M_{\oplus}$) & 15.9 & 30 & 18.75 \\
        $m_2$ ($M_{\oplus}$) & 82.6 & 200 & 178 \\  
        $P_1$ (days) & 13.3 & 17.09 & 17.09 \\
		$P_2$ (days) & 33.4 & 34.56 & 34.55 \\
        $e_1$ & $0.214$ & $0.132$ & $0.160$ \\
		$e_2$ & $0.06$ & $0.029$ & $0.0046$ \\
		$\omega_1$ ($^{\circ}$) & 240 & 193 & 292 \\
		$\omega_2$ ($^{\circ}$) & -30 & 275 & 190 \\ 
        \hline
        $m_{\star} (M_{\odot})$ & $0.76$ & $0.874$ & $0.77$ \\
		\hline
	\end{tabular}
\end{table}

\citet{dawson21} updated the orbital fit using RV observations from HARPS, FEROS and the Planet Finder Spectrograph, and expanded the TESS photometry with a ground-based transit observing campaign. This joint TTV+RV fit significantly constrained the system's properties, as can be seen in Table \ref{tab:toi216}. The authors found that TOI-216c is a warm Jupiter, TOI-216b is an eccentric warm Neptune, and that the system is actually locked in 2:1 MMR. In particular, \citet{dawson21} discussed a libration behavior of the resonant angle for a specific set of orbital conditions, and \citet{Nesvorny2022} use the libration to put constraints on primordial planet migration.

Considering the latest parameters of \citet{dawson21}, we construct a dynamical map for the TOI-216 system. The innermost planet, TOI-216b, is also the least massive one and it is the one whose parameters vary over the grid. The maps span a total integration time of 200 years. This time span encompass about 50 libration periods of the 2:1 MMR critical angle $\sigma_1$, and it is enough to reveal the perturbations that affect the system, including secular perturbations.

\begin{figure*}
    \includegraphics[width=0.49\textwidth]{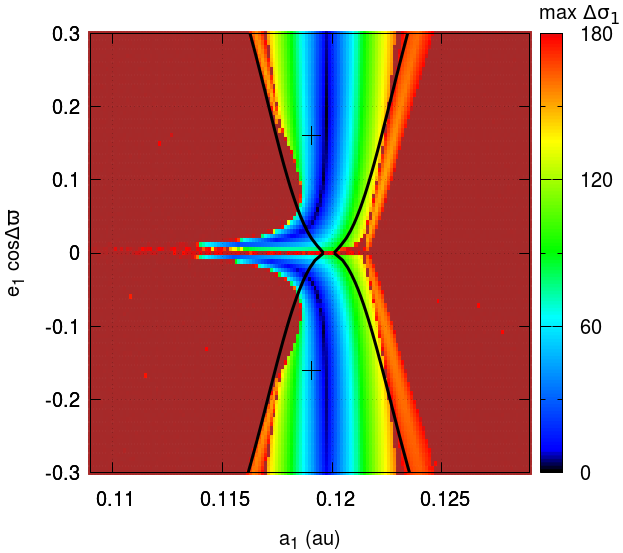}
	\includegraphics[width=0.49\textwidth]{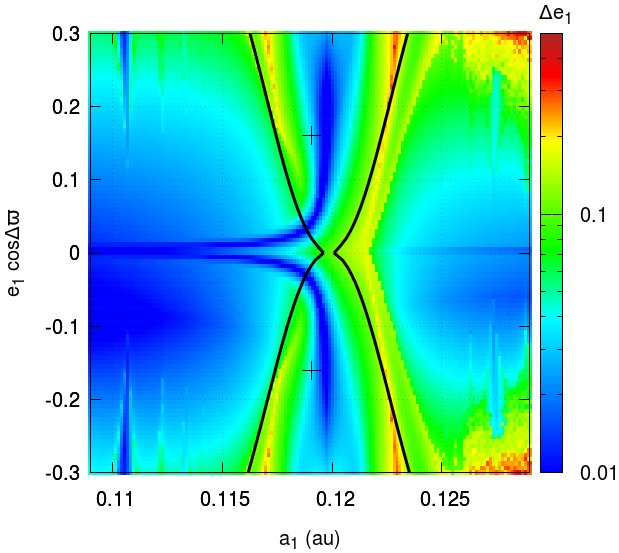}
	\includegraphics[width=0.49\textwidth]{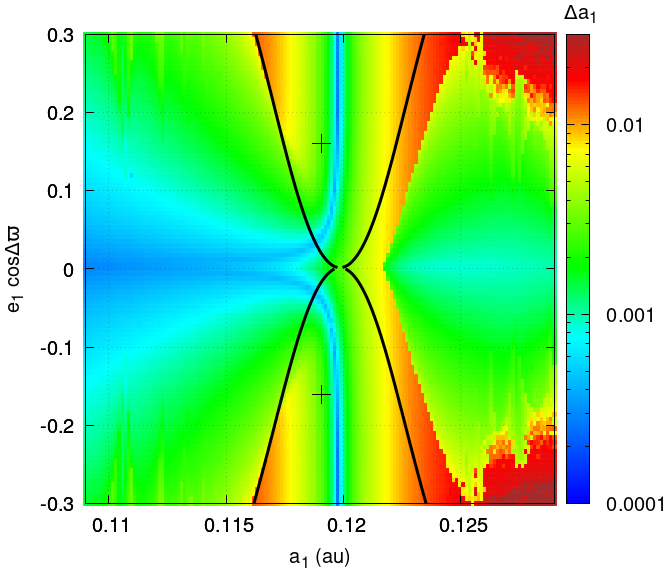}
    \includegraphics[width=0.47\textwidth]{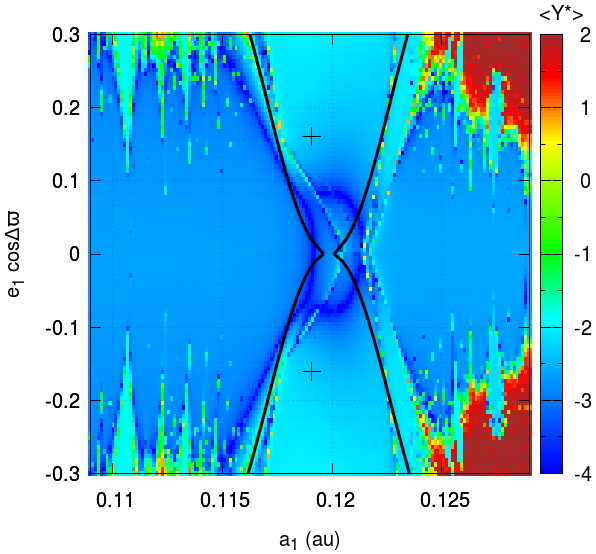}
	\caption{Dynamical maps of the TOI-216b planet in the plane $a_1, e_1 \cos \Delta \varpi$, using the orbital fit of \citet{dawson21}, i.e. $e_2=0.0046$ fixed. The maps show $\max\Delta \sigma_1$ and $\max\Delta e_1$ (top panels), and $\max\Delta a_1$ and MEGNO indicator $\langle Y^*\rangle$ (bottom panels). 
	In the top-left panel, dark blue corresponds to the deepest resonant orbits, while red represents non resonant orbits. In the other panels, dark blue represents the most regular orbits, while tones of green to red indicate increasingly irregular or chaotic motion. The semi-analytic $\delta a_1$ is overlapped as full black lines. Crosses indicate the actual values of $a_1,e_1$ for TOI-216b.
	}
    \label{fig:toi-216}
\end{figure*}

\begin{figure*}
	\includegraphics[width=0.49\textwidth]{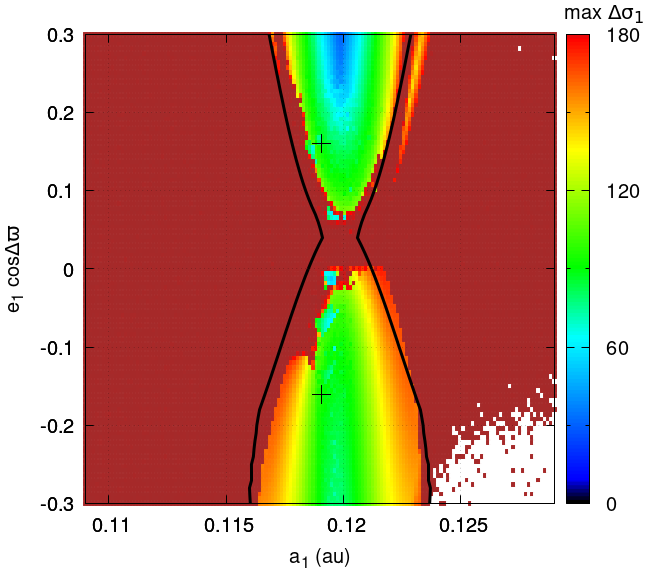}
	\includegraphics[width=0.49\textwidth]{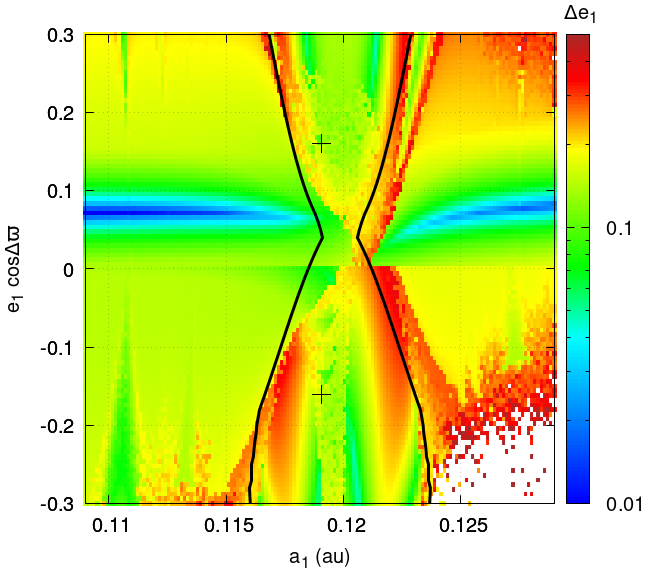}
	\includegraphics[width=0.49\textwidth]{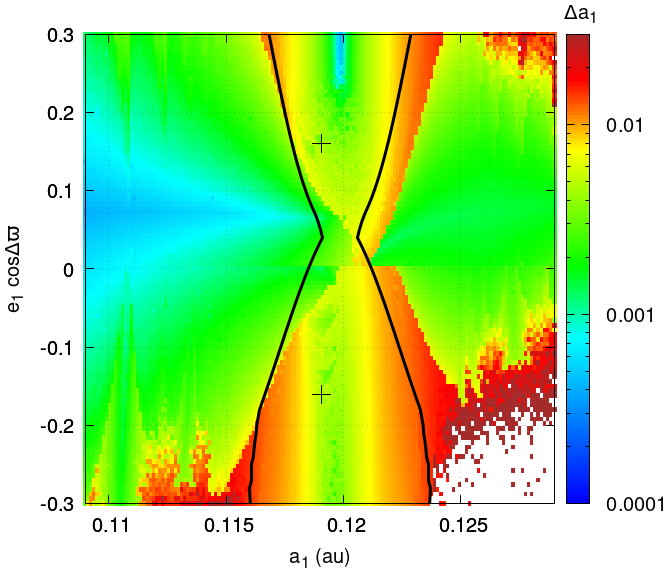}
	\includegraphics[width=0.47\textwidth]{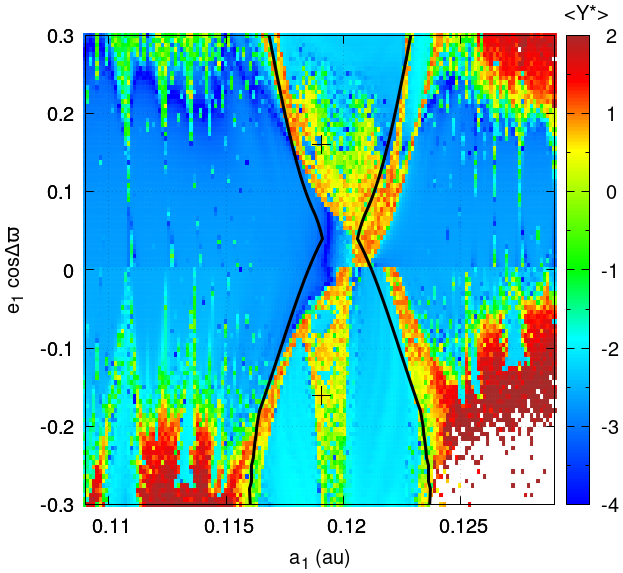}
    \caption{Similar to Fig. \ref{fig:toi-216}, but for a fictitious TOI-216 system, where the more massive planet's eccentricity is set to $e_2=0.1$. Regions in white correspond to collision orbits.}
    \label{fig:toi-216e2}
\end{figure*}

\begin{figure}
    \rotatebox{90}{%
    \begin{minipage}[b]{5cm}
    \centering
    \small\textsf{libration center (deg)}
    \end{minipage}
    }
    \begin{minipage}[b]{0.9\columnwidth}
    \centering
    \includegraphics[width=\textwidth]{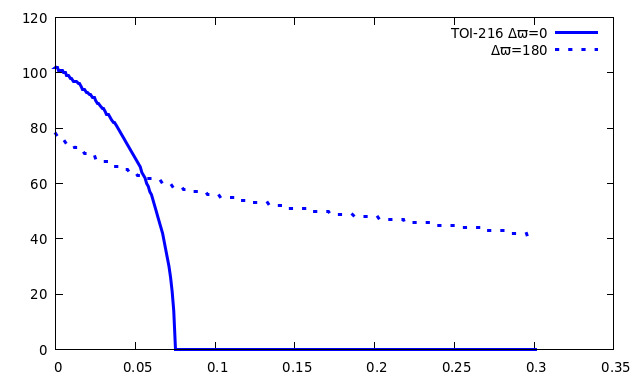}\\
    \small\textsf{eccentricity}
    \end{minipage}
    \vspace{-0.2cm}
    \caption{The locus of the stable equilibria of the resonant angle $\sigma_1=\lambda_1-2 \lambda_2+\varpi_1$ as a function of $e_1$ for the TOI-216b planet, considering a fictitious system where the eccentricity of TOI-216c is $e_2=0.1$. The plot shows two possible configurations of $\Delta\varpi$: $0^\circ$ (full line) or $180^\circ$ (dashed line). Values have been predicted by the semi-analytical model of GBG21.}
    \label{fig:equibtoi216b}
\end{figure}

\begin{figure}
   
	\includegraphics[width=0.51\textwidth]{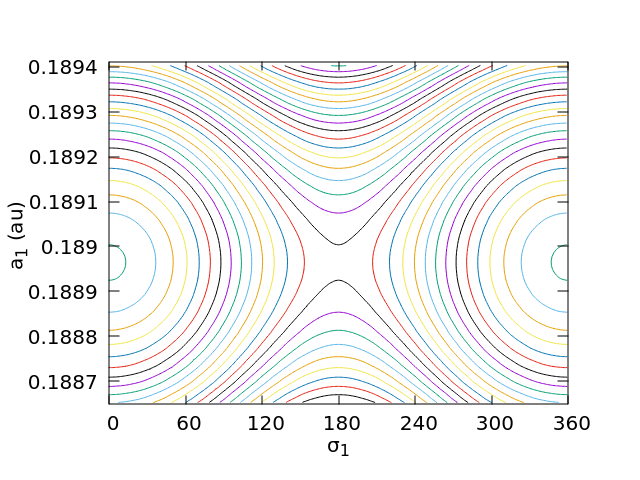}
	\includegraphics[width=0.51\textwidth]{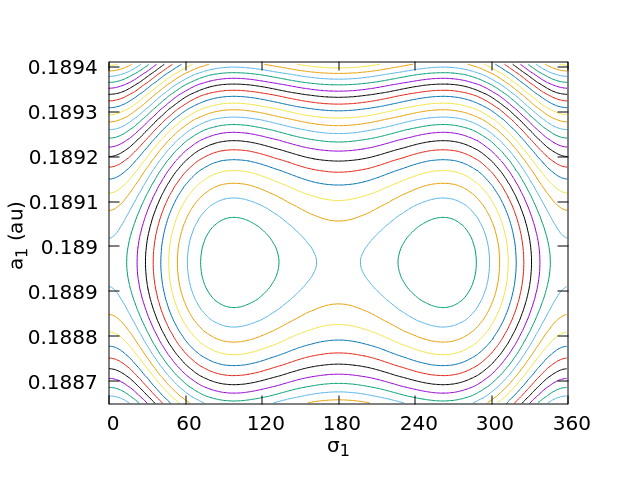}
	\includegraphics[width=0.51\textwidth]{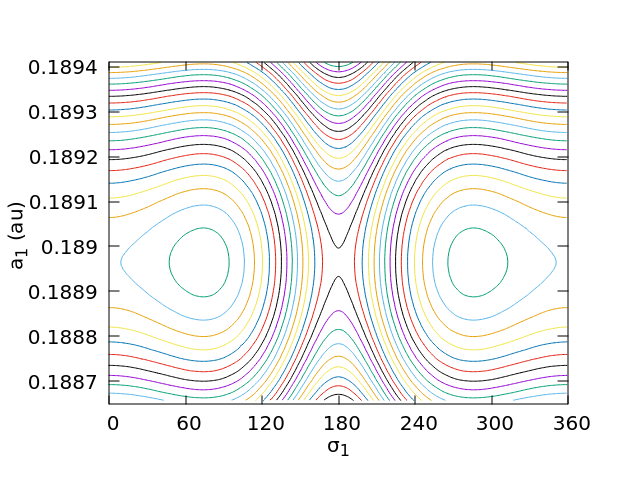}
    \caption{Level curves of constant Hamiltonian in the $a_1,\sigma_1$ space predicted by the semi-analytical model of GBG21 for the TOI-216b planet. Top: topology at $e_1=0.1$ and $\Delta\varpi=0^\circ$, with only one libration center. Middle: topology at $e_1=0.01$ and $\Delta\varpi=0^\circ$ and Bottom: $\Delta\varpi=180^\circ$, showing two libration centers, with a $180^\circ$ phase shift.}
    \label{fig:hamiltoi216b}
\end{figure}

Figure \ref{fig:toi-216} shows the dynamical maps for the nominal TOI-216 system, measuring $\max\Delta \sigma_1$ (top-left), $\max\Delta e_1$ (top-right), $\max\Delta a_1$ (bottom-left), and the MEGNO indicator $\langle Y^*\rangle$ (bottom-right). The distinction between resonant and non-resonant regimes is well visualized in the top-left panel, where the maximum amplitude of the resonant angle inside the separatrix is smaller than $180^\circ$, while outside is larger, indicating that $\sigma_1$ circulates. We recover the resonance centers along a well defined law of structure\footnote{This law can be described approximately by:
\begin{equation}
     e_1=\frac{C_1m_2}{M_{\star}\displaystyle\left(\frac{P_2}{P_1}-\beta\right)}
\end{equation}
where $C_1$ and $\beta$ are constants and $M_{\star}$ is the stellar mass \citep{Charalambous+2018}. For this particular case, $C_1=1.2$ and $\beta=2.01$.}, represented by the locus of zero amplitude libration of $\sigma_1$ (dark blue in the top-left panel), associated also to the least variations in $e_1$ and $a_1$. For very low eccentricities, the resonance centers occur for periods ratios slightly greater than 2 (up to $P_2/P_1 \sim 2.2$), or slightly smaller semi-major axes, in a regime known as pericentric paradoxical librations. 
In this regime, the oscillations of the critical angle are not associated to a separatrix, but they are small amplitude circulations around a center that is displaced from the origin.
This is the well known resonant structure that is expected in any first order MMR with $m_2/m_1>1$ and a very low eccentricity of the perturbing planet ($e_2\sim0$). It is noticeable the good correspondence between the map of critical angle amplitudes and that of maximum eccentricity variation (top panels). 

\begin{figure*}
    \centering
    \includegraphics[width=0.48\textwidth]{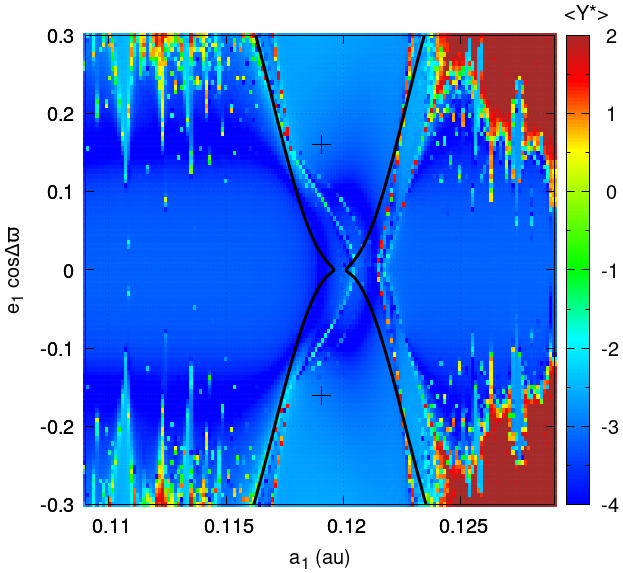}
    \includegraphics[width=0.49\textwidth]{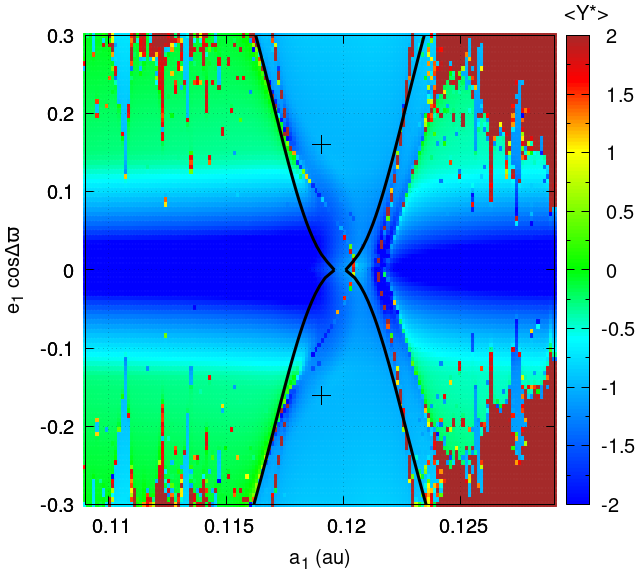}
    \caption{Maps of the MEGNO chaos indicator for the TOI-216 system in the plane $a_1, e_1 cos(\Delta \varpi$), computed over a simulation time span of 1\,000 years (left) and 10\,000 years (right). These shall be compared with the fourth panel in Fig. \ref{fig:toi-216}, that was computed over 100 years.}
    \label{fig:megnotime}
\end{figure*}

Both aligned and anti-aligned configurations of the pericenters ($\Delta\varpi$) yield the same resonant structure. 
In general, one should expect the configuration with $\Delta\varpi=180^\circ$ to be unstable. But in this system the eccentricity of the more massive planet is $e_2=0.0046$, i.e. it is almost like a circular three body problem. Therefore, unless $e_2$ increases significantly due to the perturbation of the smaller planet, the alignment or anti-alignment of the apsides is irrelevant for the stability. The fact that in this system the MEGNO map does not show any differences in spite of the integration time span, as mentioned in Sect. \ref{methods} (see also Fig. \ref{fig:megnotime}), seems to corroborate this.

The full lines in Fig. \ref{fig:toi-216} show the resonance width, $\delta a_1$, computed with the semi-analytical model of GBG21.
For $e_1\gtrsim 0.1$, we recover the V-shape of the separatrix, with a very good agreement between the numerical simulations and the semi-analytical model. 
For smaller eccentricities, however, the semi-analytical model does not fit well, because the model provides the right libration amplitude only in the regime where the separatrix exist. As the solution moves along the law of structure towards smaller eccentricities and smaller semi-major axes, into the regime of paradoxical librations, the model fails to determine which is the actual libration amplitude. This is a limitation of the model that assumes that the eccentricity $e_i$ is fixed during a single libration of $\sigma_i$, which is a bad approximation in the regime of paradoxical librations.

The crosses in Fig. \ref{fig:toi-216} indicate the fit values of $a_1,e_1$ for TOI-216b, confirming that it is in a very stable region ($\langle Y^*\rangle\sim-2$), deep inside the 2:1 MMR, with libration amplitudes smaller than $60^{\circ}$ and maximum excursions of the eccentricity of 0.04.
The errors in the fit parameters do not affect the results, since they are of the order of $7\times10^{-4}$~d for the periods and $3\times10^{-3}$ for the eccentricities. The current orbital fit also indicates that $\Delta\varpi\simeq102^{\circ}\,\!^{+50}_{-30}$.

The situation changes significantly when we consider a fictitious TOI-216 system, in which we increase the eccentricity of the more massive planet to $e_2=0.1$. The dynamical maps corresponding to this fictitious system are shown in Fig. \ref{fig:toi-216e2}. The law of structure is no longer evident and the paradoxical librations at very low eccentricities ($e_1\lesssim 0.07$) seem to disappear. Nevertheless, the semi-analytical model still reproduces well the separatrix of the libration regions that appear in the map. For $\Delta\varpi=0^\circ$, the libration region appears shifted to higher values of $e_1$. 

The behavior observed in Fig. \ref{fig:toi-216e2} arises because increasing the eccentricity $e_2$ of the more massive planet induces the bifurcation of the paradoxical librations at low values of $e_1$ into asymmetric librations of $\sigma_1$, as shown in Figs. \ref{fig:equibtoi216b} and \ref{fig:hamiltoi216b}. Using the semi-analytical model, we find that for $e_1<0.07$ and $\Delta\varpi=0^\circ$, the libration center is no longer at $0^\circ$, but two libration centers appear and shift up to $\sigma_1\sim 100^\circ$ (full line in Fig. \ref{fig:equibtoi216b}). These asymmetric librations continue to exist for $\Delta\varpi=180^\circ$ (dashed line in Fig. \ref{fig:equibtoi216b}); the apparent discontinuity between the full and dashed line in the figure has to do with the fact that $\sigma_1$ suffers a change of phase by $180^\circ$, but topologically they are the same equilibria. This is shown in Fig. \ref{fig:hamiltoi216b}, where we can see the transition from one symmetric libration center (top panel) to two asymmetric centers (bottom-left panel), and then a phase shift of the asymmetric centers (bottom-right panel).

It is worth noting that the asymmetric librations described above occur in a regime where no ACR exists. According to \citet{Beauge2006} and \citet{Voyatzis2009}, the ACR for $m_2/m_1 > 1$ only exist for $e_1 > 0.1$, and in that case the librations of both angles ($\sigma_1,\Delta\varpi$) are symmetric (around 0,0). This is in agreement with the result shown in Fig. \ref{fig:equibtoi216b}, where for $e_1 > 0.07$ the libration of $\sigma_1$ is symmetric (around 0). The asymmetric librations of $\sigma_1$ occurring for $e_1 < 0.07$ are not related to ACR and $\Delta\varpi$ is circulating in this case.

The lack of resonant librations at very low eccentricities that is evident in the left panel of Fig. \ref{fig:toi-216e2} is an artifact caused because the dynamical map has been constructed for a fixed initial value of $\sigma_1=0^\circ$, thus it fails to properly catch the asymmetric librations at very different values of $\sigma_1$. The neck of the sand clock shape in Fig. \ref{fig:toi-216e2} occurs approximately at the eccentricity where the libration point bifurcates. 
The MEGNO indicator shows that evolution inside the resonance for moderate to low eccentricities becomes more chaotic. It is curious that even in this hypothetical configuration, the current fit of TOI-216b still falls in the resonant domain, although the observational data do not support this configuration.


\newpage

\newpage

\newpage

\subsection{HD27894}\label{sec:HD27894}

HD27894 is a star of spectral type K2V with an estimated mass of $0.8~M_{\odot}$, and a metallicity of $\mathrm{[Fe/H]}=0.30\pm0.07$~dex. \citet{Moutou2005} announced HD27984b, a Jovian planet with a minimum mass of $0.62~M_{\mathrm{Jup}}$, orbiting at a semi-major axis of 0.125~au in a nearly circular orbit (eccentricity $0.049 \pm 0.008$). \citet{Kurster2015} studied the possibility of detecting an additional planet through the RV data and found that an outer Saturn-mass planet, with a period of $\approx 36$~days, provides an even better fit to the data.
Finally, \citet{Trifonov+2017} confirmed the presence of two additional planets.

\begin{table}
	\centering
	\caption{Dynamical parameters of the two innermost planets in the HD27894 system, according to \citet{Trifonov+2017} - T2017}
	\label{tab:hd27894}
    \renewcommand{\arraystretch}{1.4}
	\begin{tabular}{lr} 
		\hline
        Parameter & T2017 \\
		\hline
		$m_1$ $(M_{\oplus})$ & 211.4 \\ 
		$m_2$ $(M_{\oplus})$ & 51.5 \\ 
        $P_1$ (days) & 18.02 \\ 
		$P_2$ (days) & 36.07 \\ 
        $e_1$ & 0.047 \\ 
		$e_2$ & 0.015 \\ 
		$\omega_1$ ($^{\circ}$) & 132.2 \\ 
		$\omega_2$ ($^{\circ}$) & 44.2 \\ 
        \hline
        $m_{\star} (M_{\odot})$ & $0.8$ \\ 
        \hline
	\end{tabular}
\end{table}

We choose to study this system because it represents a good example of a system with two planets in the 2:1 MMR with a small $m_2/m_1$ mass ratio. Actually, the system can be treated as a hierarchical system, with the two innermost planets, HD27984b and HD27984c, close to the 2:1 MMR, and the third massive outer planet, HD27984d, located at a much distant orbit ($m_3\simeq5.4~M_\mathrm{Jup}$ and $a_3\simeq5.45$~au). Owing to this large separation, the secular perturbations of planet~d over the pair b-c can be neglected on the integration time span considered here. Indeed, we constructed dynamical maps including and excluding planet HD27984d, to find that there are no appreciable differences between them. Therefore, in the remaining numerical simulations, planet HD27984d was not taken into account at all. The proximity of the two innermost planets to the star may rise the question about relativistic effects on the orbits. Again, we constructed dynamical maps including and excluding relativistic corrections to find no differences between them. Following a similar reasoning, we may conclude that tidal effects do not have any influence over the analyzed time span either. Table \ref{tab:hd27894} shows the relevant parameters used to study the dynamics.

\begin{figure}
    \centering
	\includegraphics[width=0.6\columnwidth,angle=270]{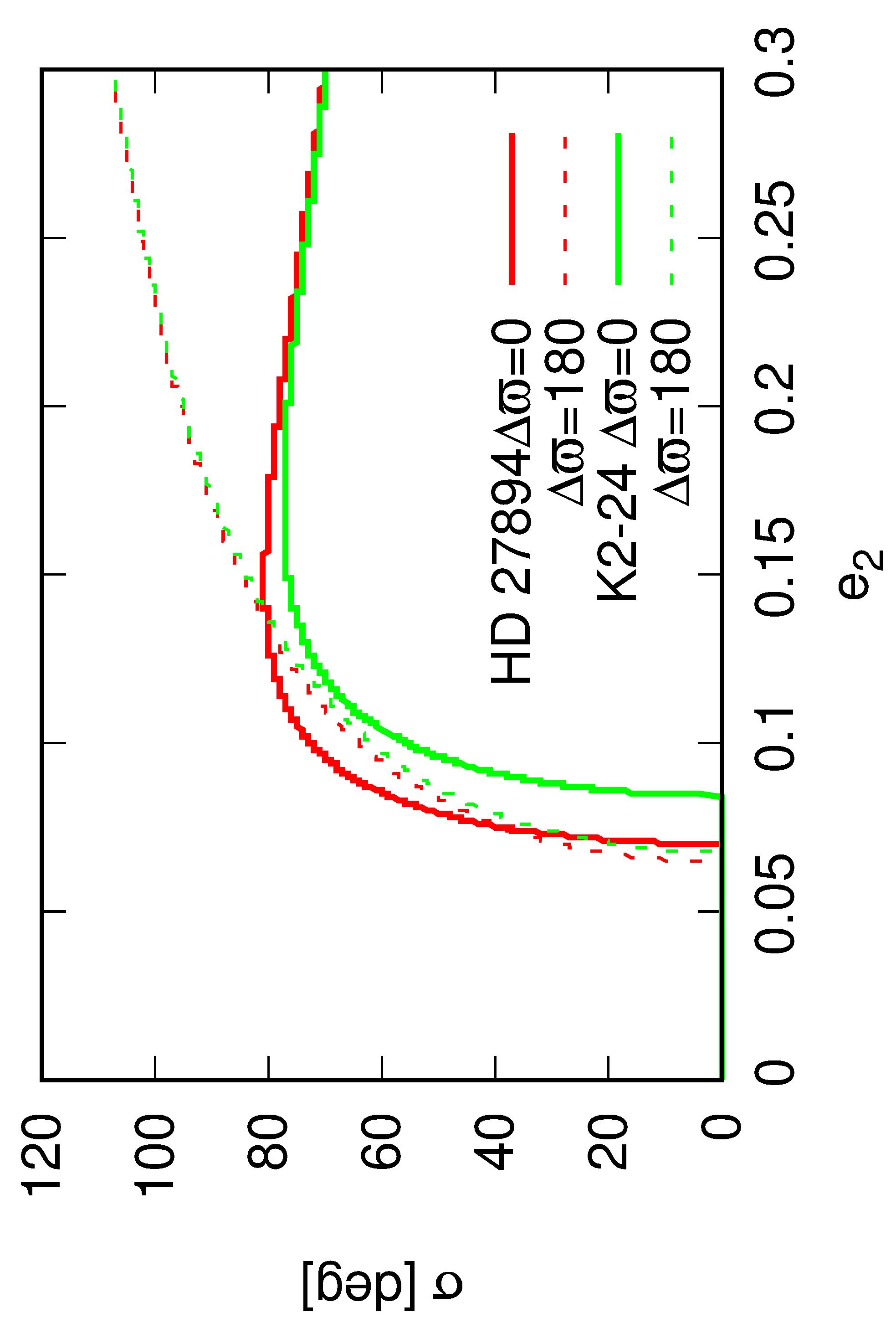}
    \caption{The locus of the stable equilibria of the resonant angle $\sigma_2$ as a function of $e_2$ for the HD27894c planet (in red; $e_1=0.047$), and the K2-24c planet (in green; $e_1=0.06$), considering two possible configurations of $\Delta\varpi=0^\circ$ (full lines) or $180^\circ$ (dashed lines). Values predicted by the semi-analytical model of GBG21.}
    \label{fig:famhd27894a}
\end{figure}

Since HD27984c is the least massive planet involved in the 2:1 MMR, we vary its orbital elements over a grid in the $a_2, e_2 \cos \Delta \varpi$ space, again assuming two initial values of $\Delta \varpi=0^\circ,180^\circ$. However, we must take into account that this system should exhibit asymmetric librations, i.e. librations of $\sigma_2$ around values different from $0^\circ ,180^\circ$, since $m_2/m_1<1$\footnote{The asymmetric librations for $m_2/m_1<1$ occur already in the framework of the restricted circular three body problem \citep{Beauge1994}.}. These asymmetric librations are predicted by the semi-analytical model of GBG21, as shown in Fig. \ref{fig:famhd27894a} (red lines). For eccentricities $\lesssim 0.07$ the resonant equilibria are at $\sigma_2=0^\circ$, but for larger values the equilibrium points bifurcate and shift to values of $\sigma_2>60^\circ$, depending on the apsidal configuration. These asymmetric centers has to be taken into account when constructing the dynamical maps. As we saw in section \ref{sec:TOI-216}, if the map grid is fixed at the same initial value of $\sigma_i$, the map will not properly capture the libration dynamics. Therefore, we use the value of $\sigma_2$ predicted by GBG21 model as a proxy for the initial value of $\sigma_2$ over the map grid, depending on the initial eccentricity. 

\begin{figure*}
	\includegraphics[width=0.49\textwidth]{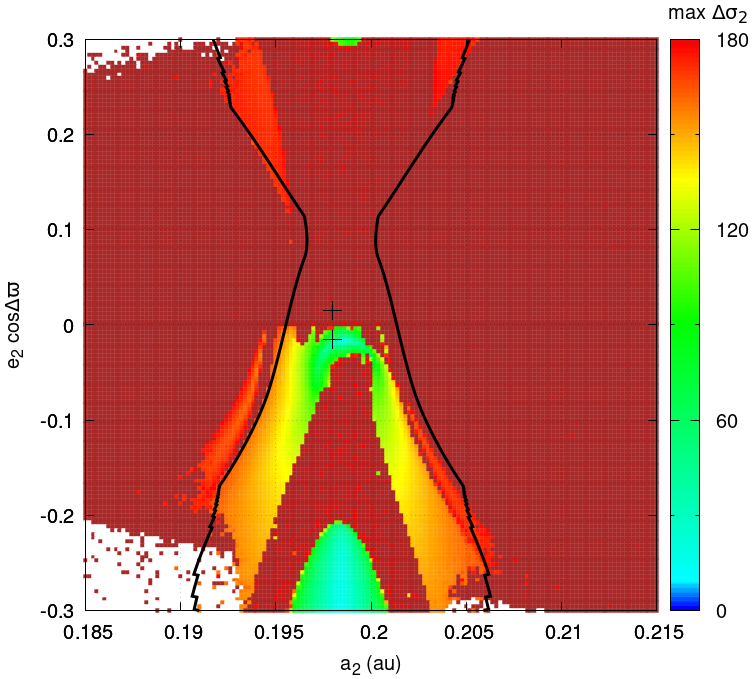}
	\includegraphics[width=0.49\textwidth]{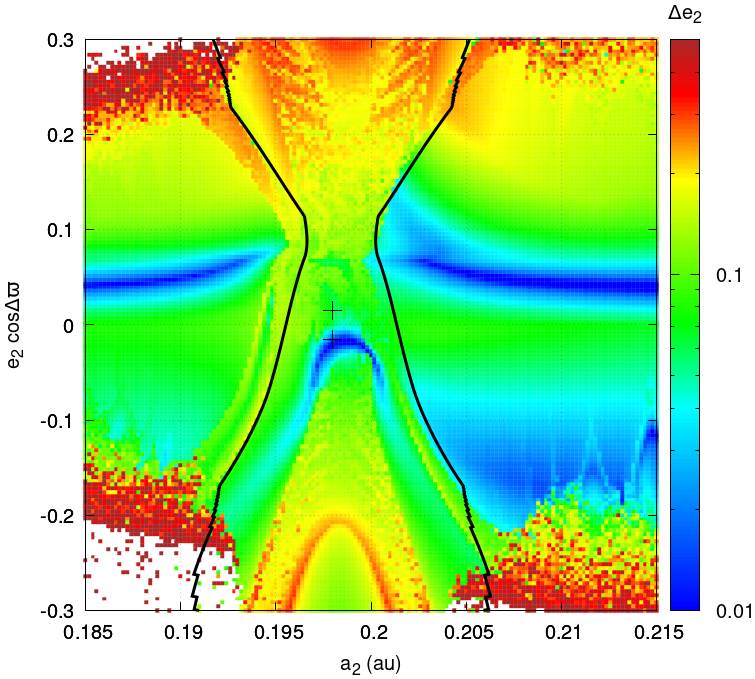}
	\includegraphics[width=0.49\textwidth]{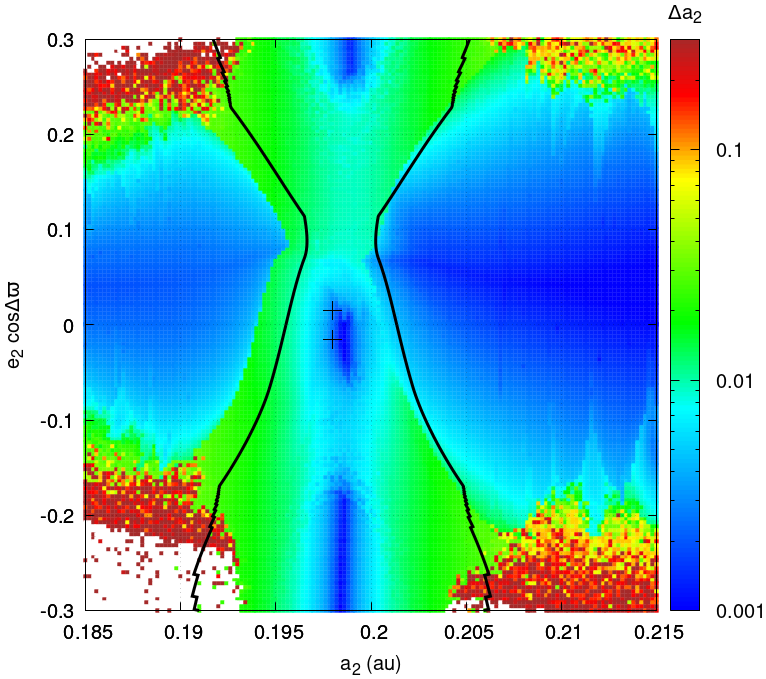}
    \includegraphics[width=0.47\textwidth]{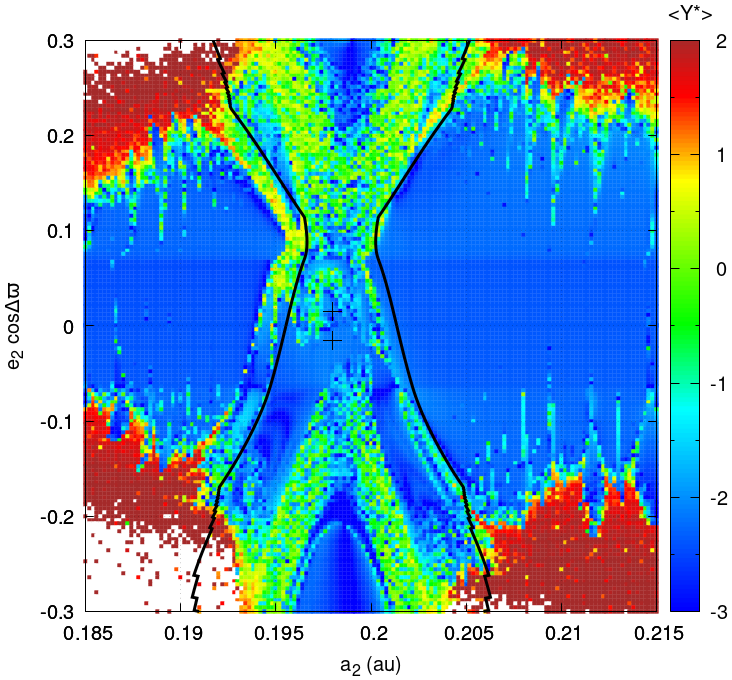}
    \caption{Dynamical maps of the HD27894c planet in the plane $a_2, e_2 \cos \Delta \varpi$, with $e_1=0.047$ fixed. The maps show $\max\Delta \sigma_2$ and $\max\Delta e_2$ (top panels), and $\max\Delta a_2$ and the MEGNO indicator $\langle Y^*\rangle$ (bottom panels). The color coding is the same as in Figs. \ref{fig:toi-216}-\ref{fig:toi-216e2}. White regions correspond to collision orbits. The maps have been constructed using a grid where $\sigma_2$ varies with $e_2$ as shown in Fig. \ref{fig:famhd27894a}. 
    The semi-analytic $\delta a_2$ is overlapped as full black lines. Crosses indicate the actual values of $a_2,e_2$ for HD27894c.}
    \label{fig:hd27894}
\end{figure*}

\begin{figure*}
	\includegraphics[width=0.43\textwidth]{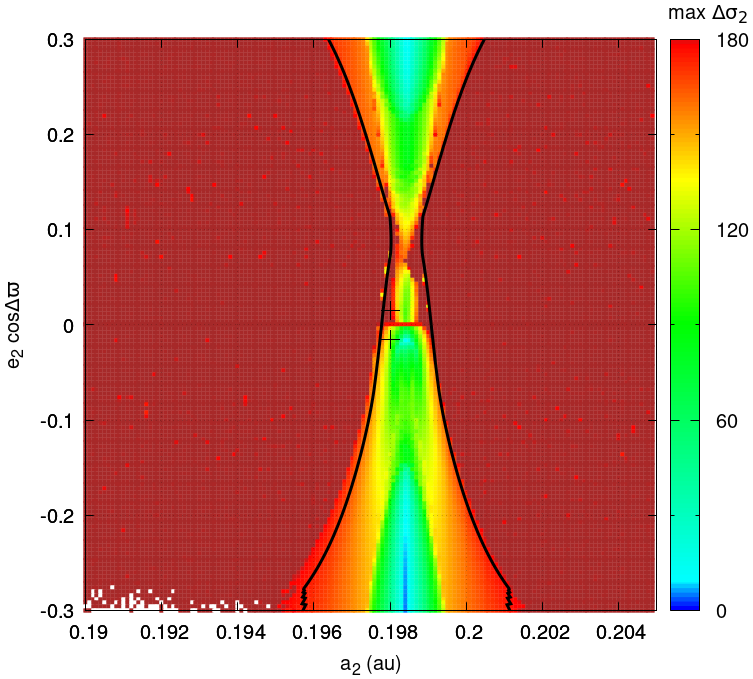}
    \includegraphics[width=0.43\textwidth]{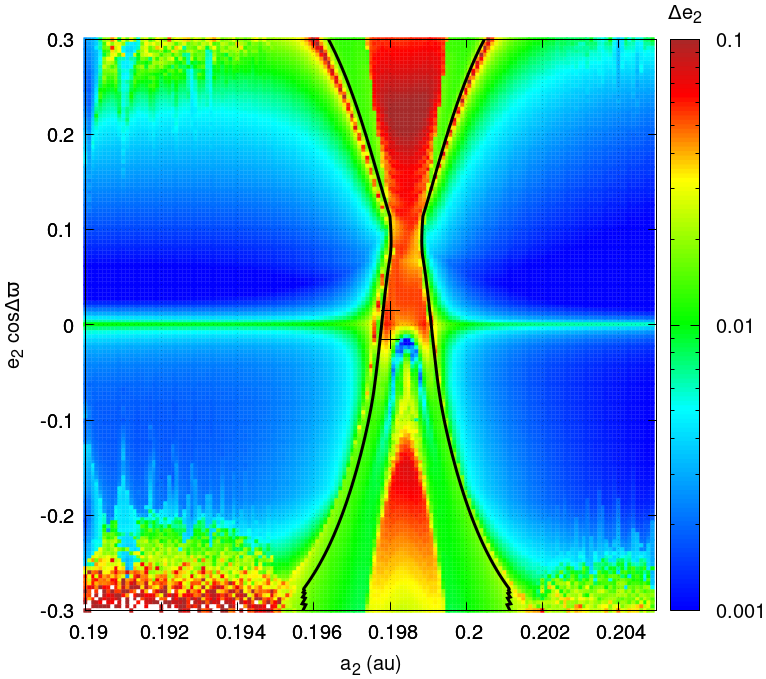}\\
    \includegraphics[width=0.43\textwidth]{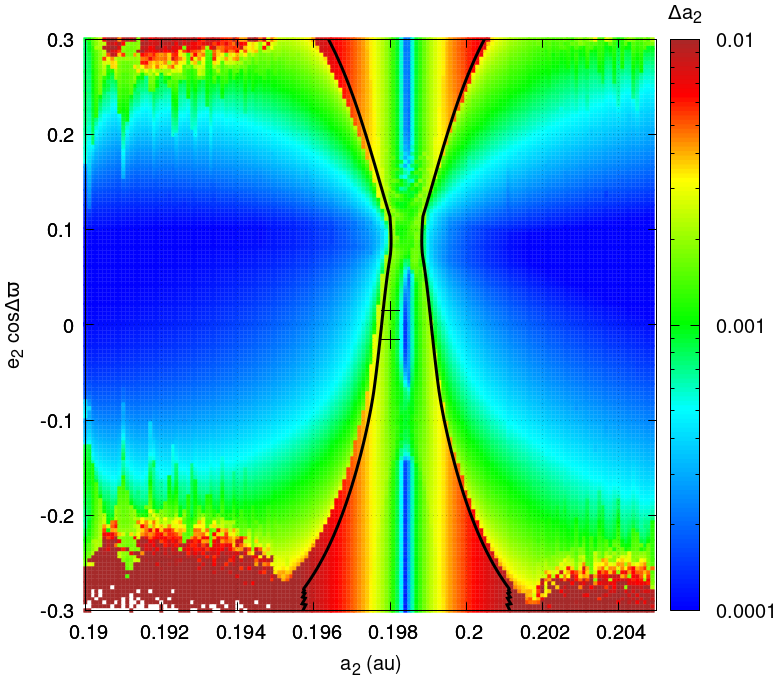}
    \includegraphics[width=0.4\textwidth]{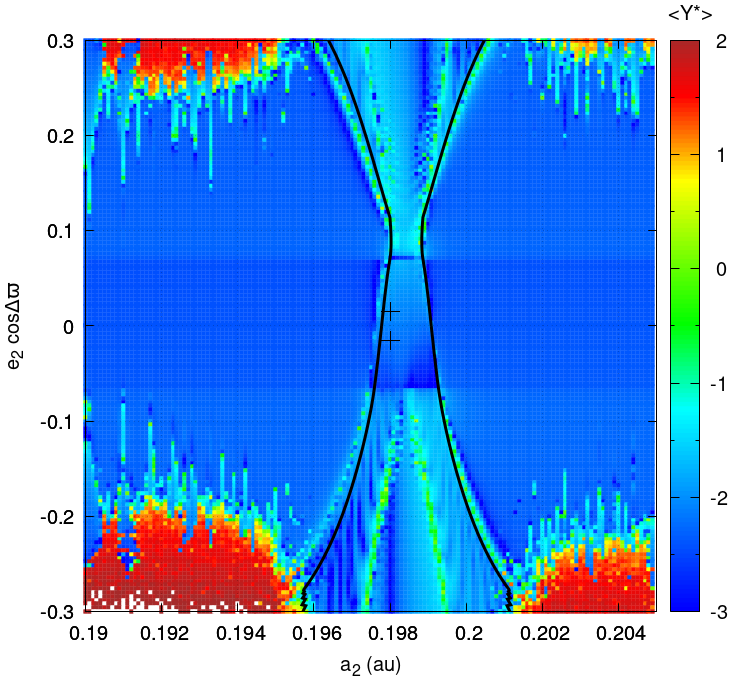}
    \caption{Same as Fig. \ref{fig:hd27894}, but for a fictitious HD27894 system, with the planetary masses $m_1,m_2$ scaled down by a factor of 20.}
    \label{fig:hd27894varmas}
\end{figure*}

\begin{figure*}
    \includegraphics[width=0.49\textwidth]{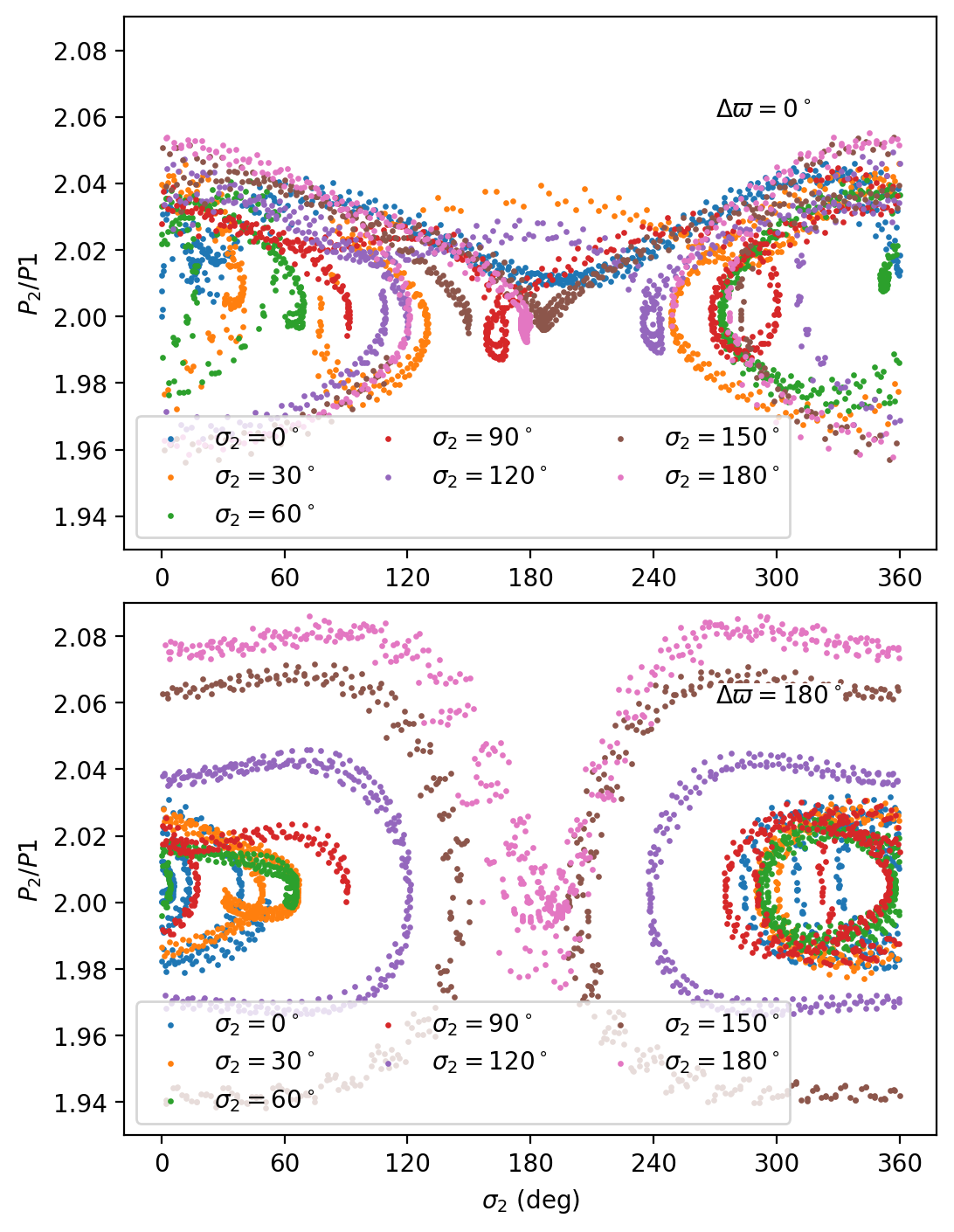}
    \includegraphics[width=0.49\textwidth]{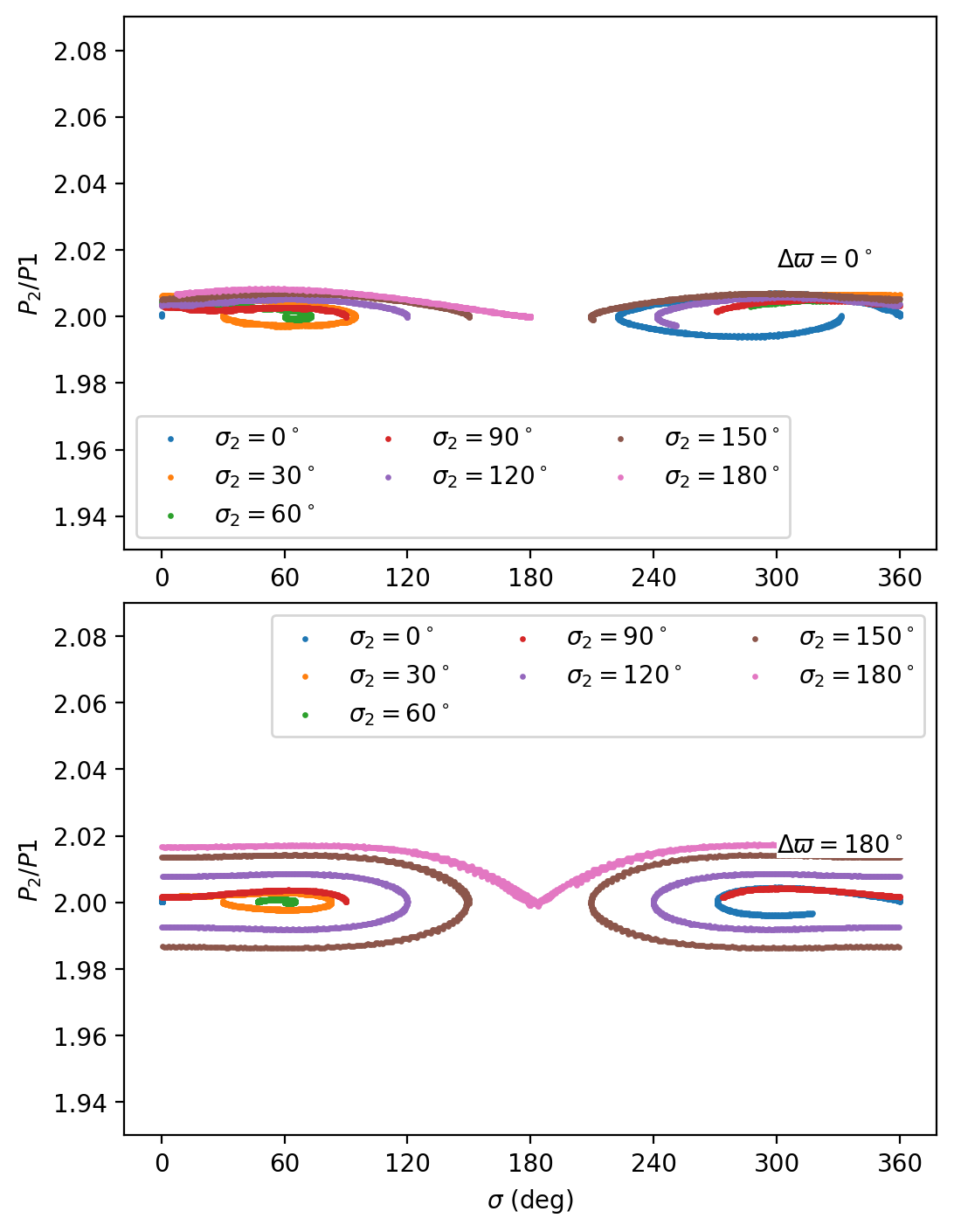}
    \caption{Short numerical integration of the orbit of HD27894c, for different initial values of $\sigma_2$ and $a_2$ (or $P_2/P_1$), assuming $e_1=0.047$ and $e_2=0.015$ fixed. Top panels correspond to initial $\Delta\varpi=0^\circ$. Bottom panels correspond to initial $\Delta\varpi=180^\circ$. Left panels consider the actual masses of the planets, as in Fig. \ref{fig:hd27894}; the integration time span is 10 years. Right panels consider the scaled down masses, as in Fig. \ref{fig:hd27894varmas}; the integration time span in 20 years.}
\label{fig:famhd27894b}
\end{figure*}

Figure \ref{fig:hd27894} shows the dynamical maps of HD27984c. The full black lines represent the semi-analytical resonance width, and the cross indicates the current values of $a_2,e_2$ of the planet. 
The errors in the fit parameters do not significantly affect the location of the planet in the maps, since they are of the order of at most 0.001 au for the semi-major axes and 0.02 for the eccentricities.
The most relevant feature is that, for initial $\Delta\varpi=0^\circ$, there is no resonant libration region, not even inside the predicted semi-analytic boundaries (and even if the map grid is following the locus of the asymmetric librations). Librations only occur for values of $\Delta\varpi=180^\circ$, where the semi-analytic width correctly encompasses the resonant domain. The map of MEGNO also indicates that, for aligned pericenters, there is a chaotic onset of orbits, and regular motion mostly occurs for anti-aligned pericenters.  We conclude that only under the condition $\Delta\varpi=180^\circ$, the actual planets pair would be locked into a resonant configuration. The orbital fit, however, provides $\Delta\varpi\simeq88^{\circ}\,\!^{+10}_{-20}$, indicating that the system could not be evolving in the MMR. This is in agreement with the results of \citet{Trifonov+2017}, who performed long term simulations of the system and did not detect any resonant behavior over 10 My of evolution.

On the other hand, Fig. \ref{fig:hd27894varmas} shows the dynamical maps for a fictitious HD27984 system, where the masses of both planets b and c have been scaled down by a factor of 20. The resonance width in this case is narrower, as expected, but the libration region clearly appears for both values of $\Delta\varpi$. The resonant domain shows a sand clock shape, with the neck displaced to high eccentricities (cf. Fig. \ref{fig:deltaa}).
It is worth noting that this particular configuration is not supported by the current observations of the system, but its analysis will help to better interpret the maps of the actual system.

The differences between Figs. \ref{fig:hd27894} and \ref{fig:hd27894varmas} can be understood by analyzing the behavior of the resonant angle $\sigma_2$ over a short time scale. Figure \ref{fig:famhd27894b} shows this behavior in the $P_2/P_1,\sigma_2$ space considering the actual masses of the system (left panels) over a time span of 10 years, and the scaled-down-mass system (right panels) over a time span of 20 years. For the actual masses and $\Delta\varpi=0^\circ$, the mutual perturbations between the planets quickly destroy any resonant behavior, and librations only survive when $\Delta\varpi=180^\circ$. Moreover, the MEGNO indicator in Fig. \ref{fig:hd27894} shows that the actual planetary masses lead to chaotic behavior inside the resonance, where few islands of regular motion are found. On the other hand, for the scaled down masses, the mutual perturbations are not so strong, and the resonant librations survive for any configuration of $\Delta\varpi$. 

\newpage
\subsection{K2-24}\label{sec:K2-24}

\citet{Petigura+2016} reported the discovery of two sub-Saturn planets orbiting K2-24, a bright ($\mathrm{V}=11.3$) metal-rich ($\mathrm{[Fe/H]}=0.42\pm0.04$~dex) G3 dwarf, in the K2 Campaign 2 field. These authors provided mass measurements of $m_1 = 23.2~M_\oplus$ and $m_2 = 31~M_\oplus$ based on Keck/HIRES RVs spanning one observing season. They also predicted TTV amplitudes of several hours, based on the proximity of the planets to the 2:1 MMR. The two planets have orbital periods of 20.9~days and 42.4~days, lying only 1\% outside the nominal 2:1 MMR.  Mass constraints were updated by \citet{Dai+2016} using additional RV data, obtaining $m_1 = 19.8~M_\oplus$ and $m_2 = 26~M_\oplus$. A more recent study by \citet{Petigura+2018} resulted in orbital fits that show low, but non-zero, eccentricities of $e_1 \simeq e_2 \simeq 0.06\pm0.01$, and masses of $m_1=19\pm2~M_\oplus$ and $m_2=15\pm2~M_\oplus$. K2-24c is only 20\% less massive than K2-24b, despite being 40\% larger; their large sizes and low masses imply larger envelopes \citep{Petigura+2018}. The variety of orbital fits is reflected in Table \ref{tab:K2-24}.

\begin{table}
	\centering
	\caption{Dynamical parameters of the K2-24 system, according to different fits: \citet{Petigura+2016} - P2016, \citet{Dai+2016} - D2016, and  \citet{Petigura+2018} - P2018.}
    \label{tab:K2-24}
    \renewcommand{\arraystretch}{1.4}
	\begin{tabular}{lrrr}
        \hline
        Parameter & P2016 & D2016 & P2018 \\
		\hline
		$m_1$ $(M_{\oplus})$ & 23.2 & 19.8 & 19.0 \\
		$m_2$ $(M_{\oplus})$ & 31.0 & 26.0 & 15.4 \\
		$P_1$ (days) & 20.89 & - & 20.89 \\
		$P_2$ (days) & 42.36 & - & 42.34 \\
		$e_1$ & $0.24$ & $0.24$ & $0.06$ \\
		$e_2$ & $<0.39$ & $<0.58$ & $<0.07$ \\
		$\omega_1$ ($^{\circ}$) & - & - & - \\
		$\omega_2$ ($^{\circ}$) & - & - & - \\
        \hline
        $m_{\star} (M_{\odot})$ & $1.12$ & $1.12$ & $1.07$ \\
		\hline
	\end{tabular}
\end{table}

\begin{figure*}
    \centering
	\includegraphics[width=0.4\textwidth]{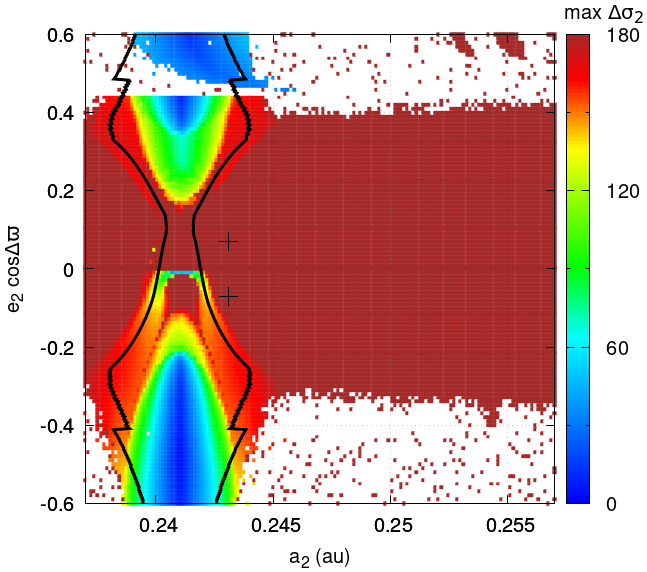}
	\includegraphics[width=0.4\textwidth]{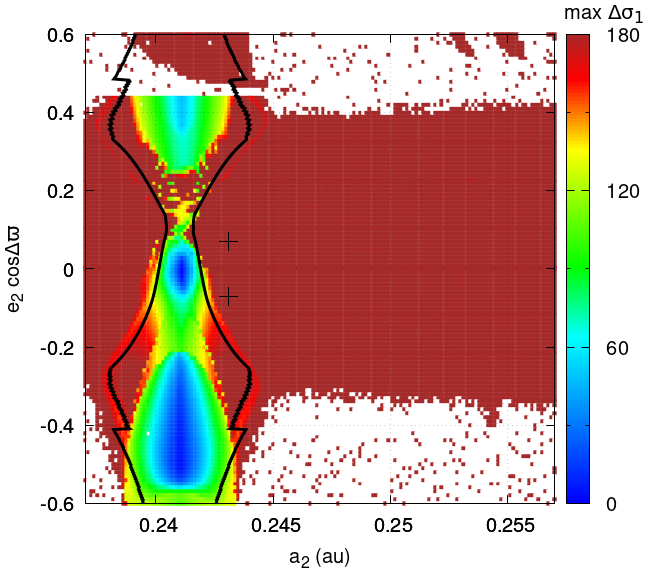}
	\includegraphics[width=0.4\textwidth]{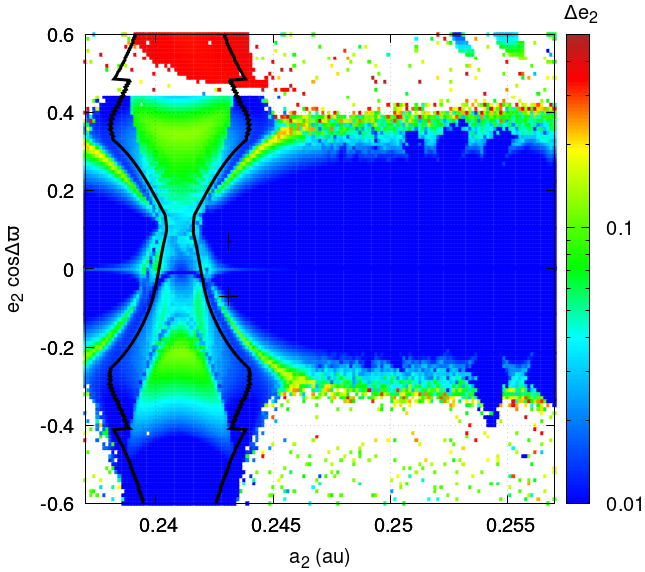}
	\includegraphics[width=0.4\textwidth]{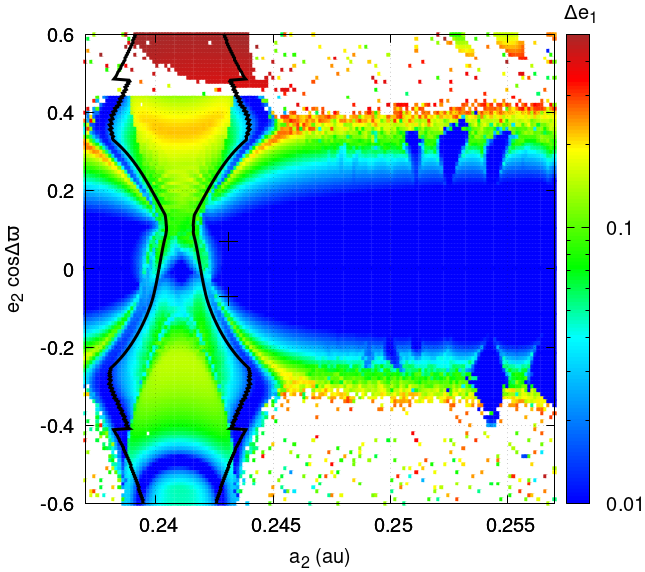}
	\includegraphics[width=0.4\textwidth]{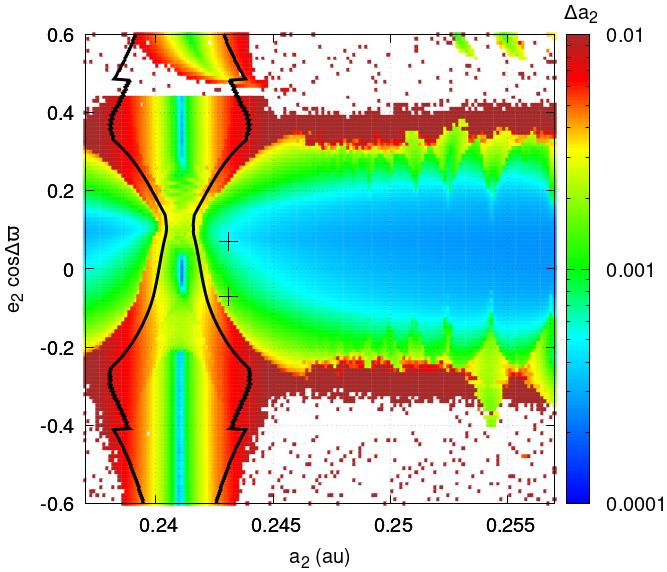}
	\includegraphics[width=0.4\textwidth]{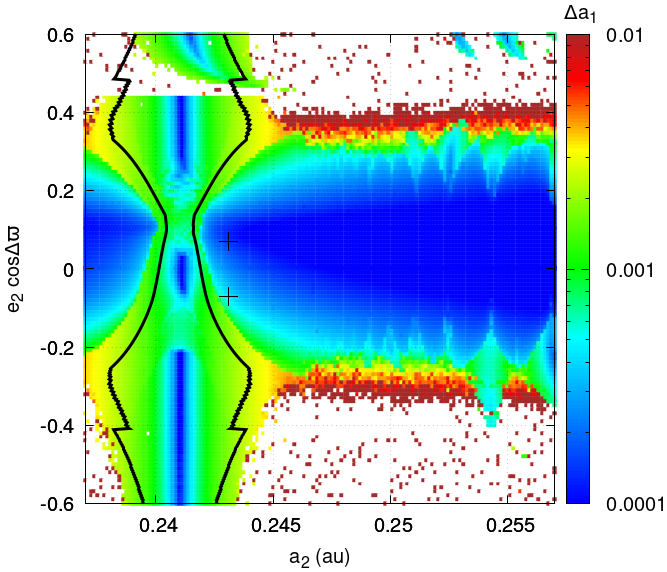}
    \caption{Dynamical maps of the K2-24 system in the plane $a_2, e_2 \cos \Delta \varpi$, considering the orbital fit of \citet{Petigura+2018}, i.e. $e_1=0.06$ fixed. The maps show, from top to bottom, the $\max\Delta \sigma_i$,  $\max\Delta e_i$, and $\max\Delta a_i$ for each planet. Left panels display the variations of K2-24c, while right panels display the variations of K2-24b, both computed over the same grid (i.e. varying K2-24c). The value of $\sigma_2$ varies with $e_2$ depending on the locus of the asymmetric libration centers (cf. Fig. \ref{fig:famhd27894a}). The value of $\sigma_1$ also varies accordingly. The color coding is the same as in Fig. \ref{fig:toi-216}. White regions correspond to collision orbits. The semi-analytic $\delta a_2$ is overlapped as full black lines. Crosses indicate the actual values of $a_2,e_2$ for K2-24c.}
    \label{fig:k2-24}
\end{figure*}

We choose to study this system because it represents a good example of a 2:1 MMR system with $m_2/m_1\sim 1$ mass ratio. Nevertheless, since the exterior planet is slightly less massive than the interior one, this system should also display asymmetric librations in a similar way as the HD27894 system. Therefore, we construct dynamical maps considering a grid in the $a_2, e_2\cos\Delta\varpi$ space, i.e. varying the parameters of K2-24c while keeping planet b fixed. The values of $\sigma_2$ vary over the grid following the predicted values in Fig. \ref{fig:famhd27894a} (green lines), in order to account for the presence of the asymmetric librations. It is worth noting that $\sigma_1$ will also display asymmetric librations, and these have been taken into account when setting the initial conditions of the simulations over the grid.

Since the planets have similar masses, it is relevant in this case to analyze the variations of the orbital elements of both planets. The dynamical maps are shown in Fig. \ref{fig:k2-24}. The left panels display the variations in the orbit of K2-24c, while the right panels display the variations in the orbit of K2-24b. In spite of the small eccentricities of both planets, the resonant portraits resemble again a sand clock shape, with the neck shifted to about $e_2\sim0.1$. The semi-analytical model predicts a resonance width in very good agreement to the numerical simulations. 

The planets would be interacting in crossing orbits for values of $e_2\gtrsim 0.3$-0.4, depending on $\Delta\varpi$. The maximum width of the resonant domain occurs approximately at the same eccentricity of the crossing orbits. The planets would be protected form close encounters only inside the resonance. For $\Delta\varpi=180^\circ$, this protection mechanism seems to extend to very high eccentricities. On the other hand, for $\Delta\varpi=0^\circ$, there is a region at $e_2\sim0.5$ where the protection mechanism inside the resonance fails, but it is present again at higher eccentricities where only $\sigma_2$ librates.

The region of the sand clock neck does not display librations of $\sigma_2$, but it does display of $\sigma_1$. We also note that the variations in $e_1,e_2$ may be significant there ($\sim0.05$-0.1). This region is associated to the bifurcation of $\sigma_2$, that passes from a symmetric libration at $e_2\lesssim0.08$ to asymmetric librations at higher eccentricities. The fact of having planets with similar masses may cause perturbations that contribute to destroy the librations of $\sigma_2$, in the same way as in the HD27894 case (cf. Fig. \ref{fig:famhd27894b}). 

The actual system, represented by crosses in Fig. \ref{fig:k2-24} lies close to the resonance border but well outside the MMR.
The errors in the fit parameters do not affect this result, since they are of the order of $10^{-3}$ d for the periods and the eccentricities are bounded to 0.1. It is worth noting that the pericenter longitudes are not constrained from the observational data.


\newpage
\subsection{The sand clock}
\label{sec:sand-clock}

Along the previous sections, we have verified that whenever the eccentricity of the more massive planet is equal or higher than $\sim0.05$, the shape of the resonant domain in the $a,e$ space resembles that of a sand clock, instead of the classical V-shape. 

At variance with the V-shape, that is symmetric with respect to $\Delta\varpi=0^\circ,180^\circ$, the sand clock is not symmetric and the neck of the clock, i.e. the place where the width $\delta a$ is minimum, always occur for $\Delta\varpi=0^\circ$. This neck is related to the occurrence of bifurcations of the $\sigma$ libration centers. For the outer planet, $\sigma_2$ passes from a symmetric libration at low $e_2$ to asymmetric librations at high $e_2$. For the inner planet, $\sigma_1$ passes from a symmetric libration at high $e_1$ to asymmetric librations at low $e_1$. For $\Delta\varpi=180^\circ$ the shape of the resonant domain appears as a continuation of the lower bulb of the sand clock.

\begin{figure}
	\includegraphics[width=0.35\textwidth,angle=270]{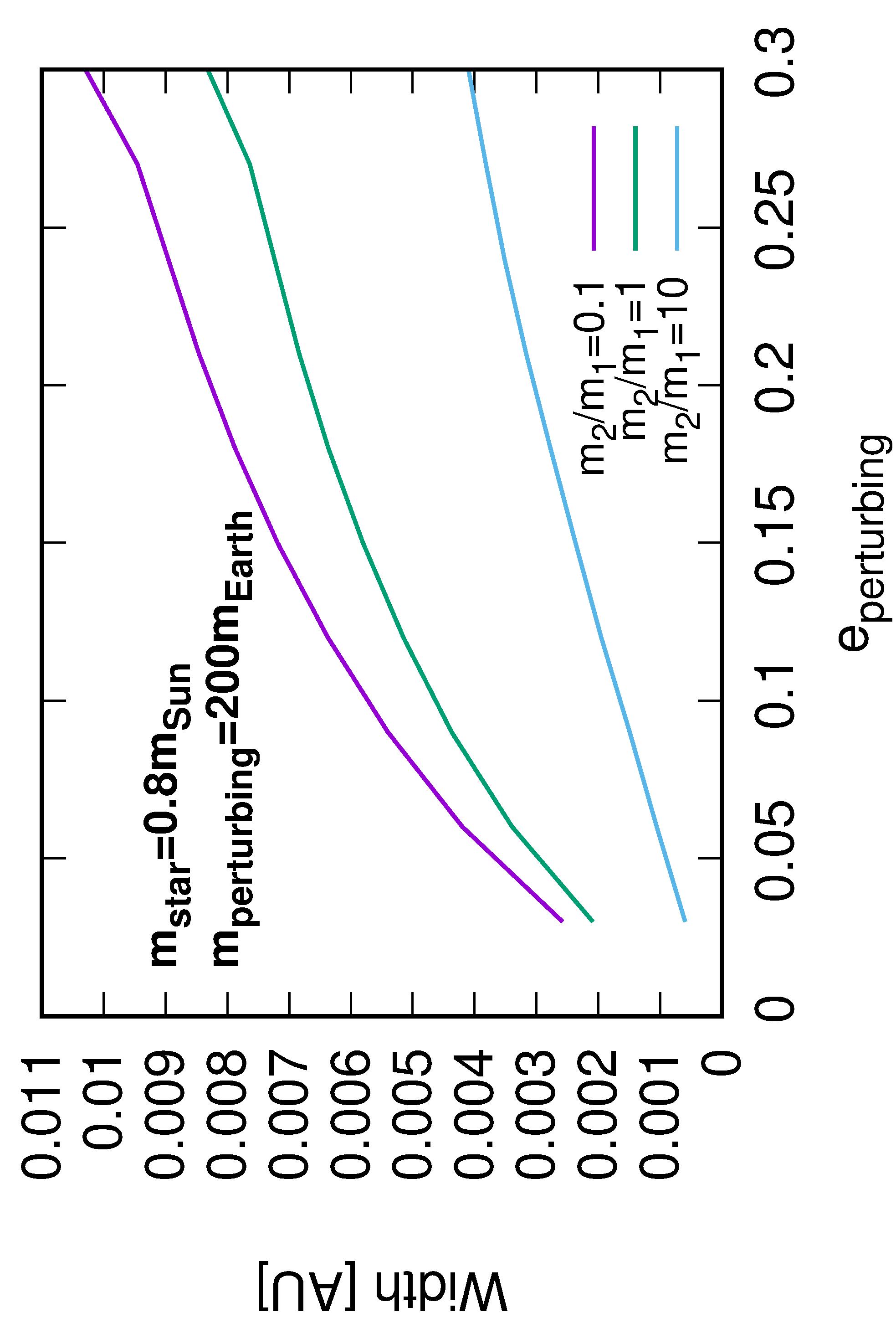}
	\includegraphics[width=0.35\textwidth,angle=270]{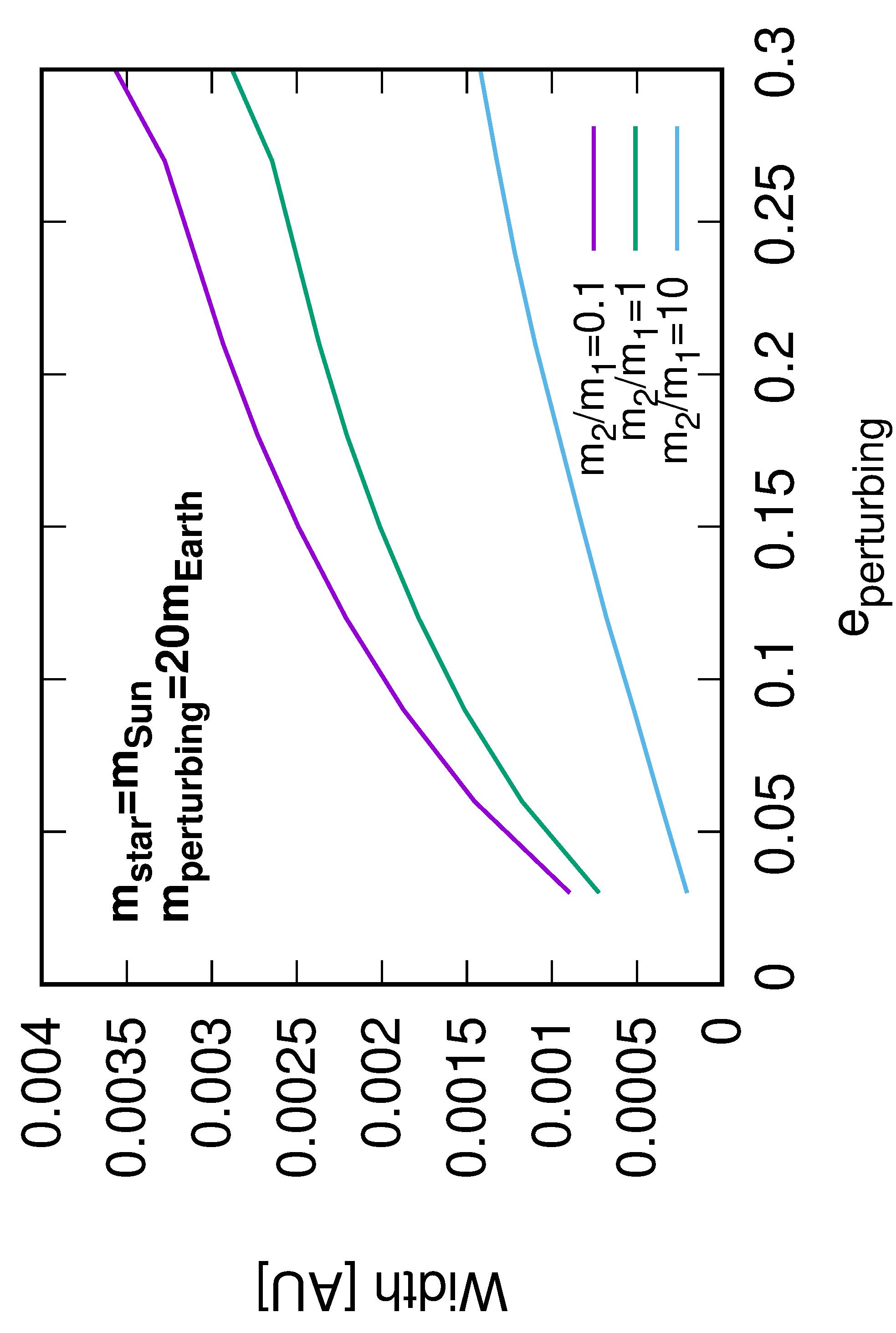}
    \caption{The width of the sand clock neck as a function of the eccentricity of the more massive (perturbing) planet. The panels correspond to different masses of the perturbing planet, and the color lines represent different mass ratios. For $m_2/m_1=1$, $e_\mathrm{perturbing}$ is assumed to be $e_1$. Values computed with the semi-analytical model of GBG21}
    \label{fig:larguras}
\end{figure}

\begin{figure}
	\includegraphics[width=0.35\textwidth,angle=270]{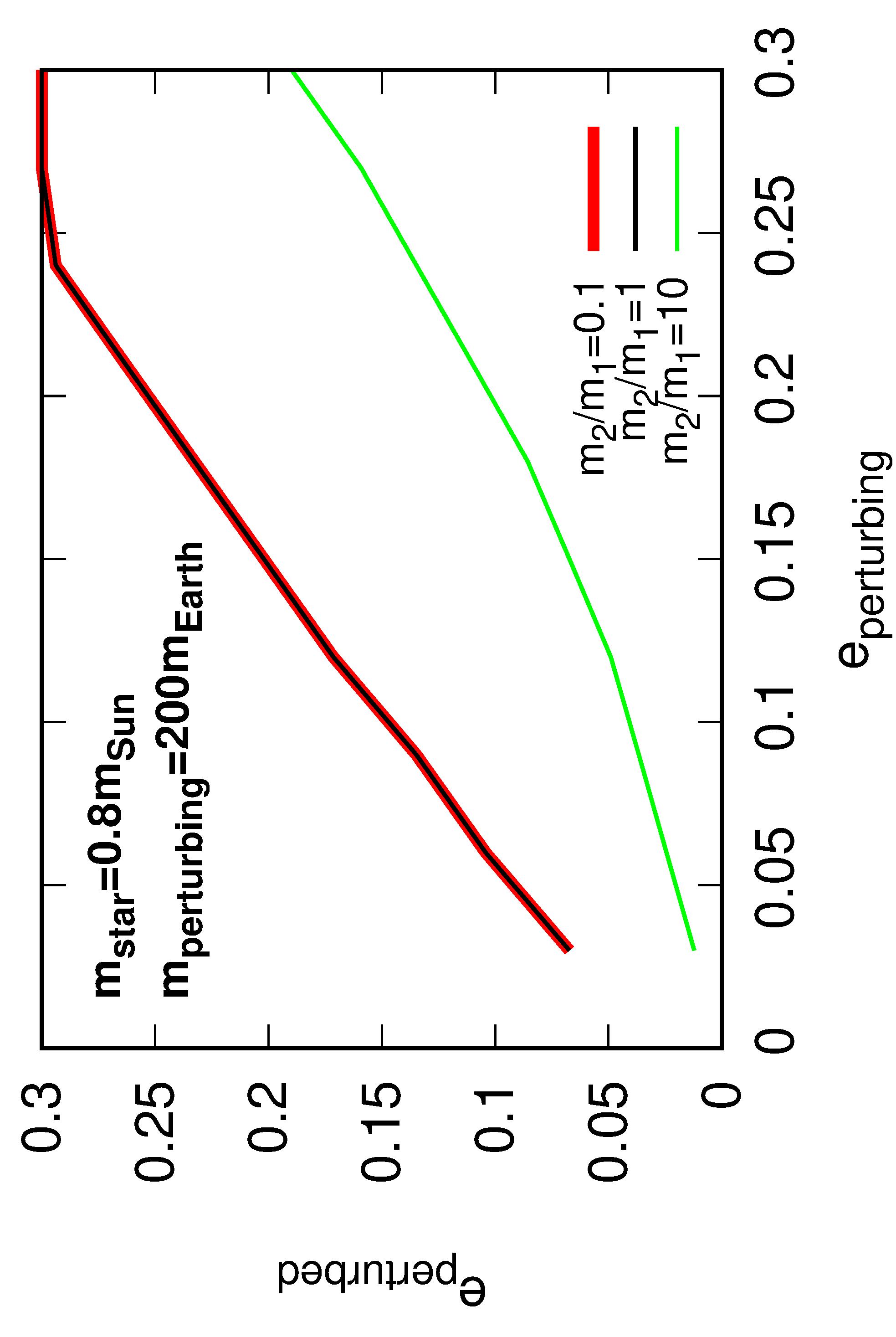}
	\includegraphics[width=0.35\textwidth,angle=270]{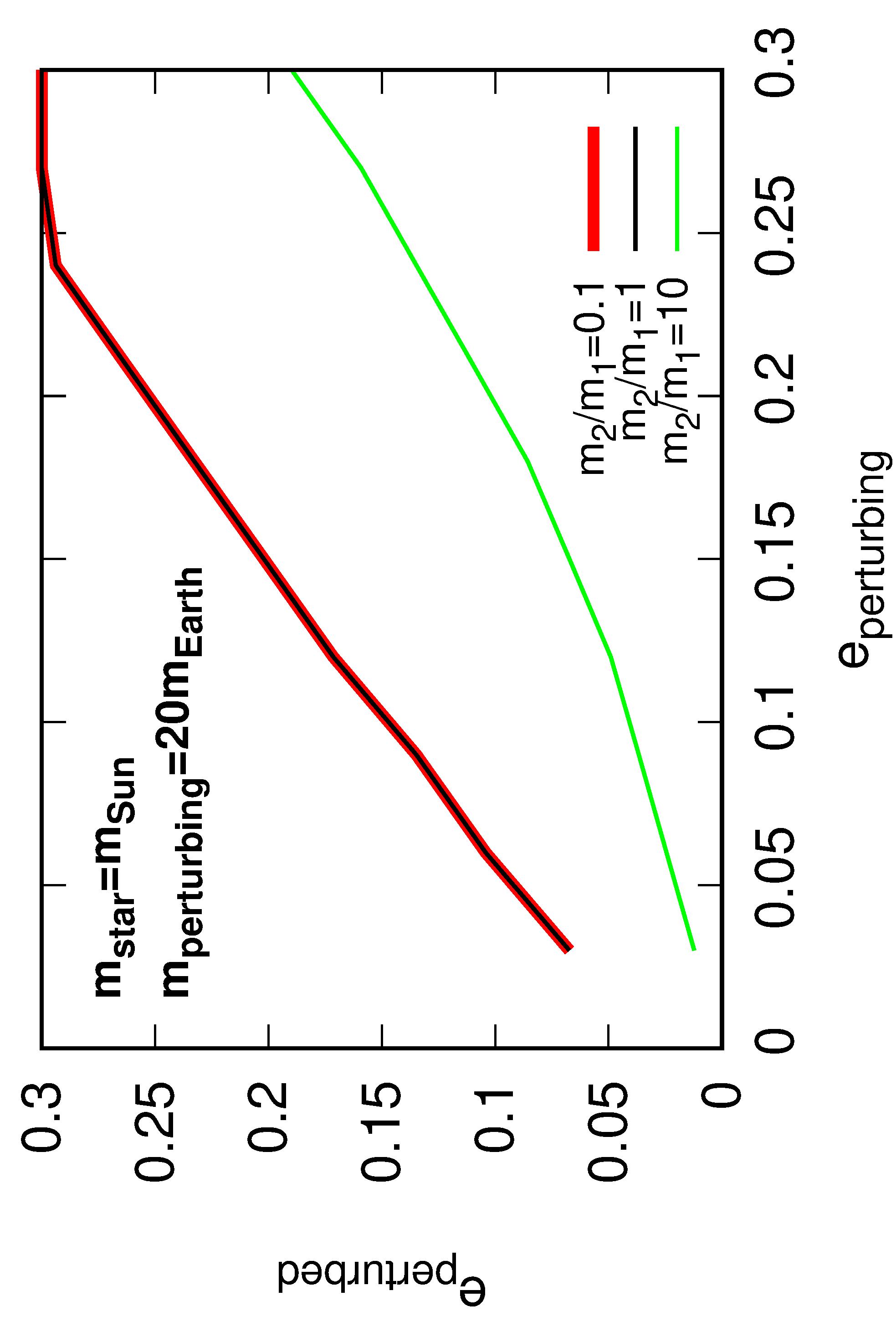}
    \caption{Similar as Fig. \ref{fig:larguras}, but for the locus of the sand clock neck in the eccentricity of the less massive (perturbed) planet. Note that the black and red lines overlap each other.}
    \label{fig:ecces}
\end{figure}

We may try to characterize the sand clock shape using two quantities: (i) the width of the neck, and (ii) the eccentricity of the less massive (perturbed) planet at which the neck occurs. Figure \ref{fig:larguras} shows the first quantity as a function of the eccentricity of the more massive (perturbing) planet. The different line colors correspond to different mass ratios, and we have considered two different scenarios: one with a large mass of the more massive planet (left panel), similar to the TOI-216 or HD27894 systems ($\sim 200\,M_\oplus$), and another with a small mass of the more massive planet (right panel), similar to the K2-24 system ($\sim 20\,M_\oplus$). 

We can see that, in both cases, the neck width increases with the eccentricity of the more massive planet, and tends to zero for a circular orbit (as in the case of TOI-216b, cf. Fig. \ref{fig:toi-216}). The neck width decreases for increasing mass ratios $m_2/m_1$. It also decreases with the mass of the more massive planet, and if we take the neck width as a proxy of the overall resonance width, this latter behavior is already expected\footnote{Recall that in the restricted circular three body problem, the resonance width scales as $\sqrt{m_\mathrm{pert}}$.}.

On the other hand, Fig. \ref{fig:ecces} shows the eccentricity locus of the neck as a function of the eccentricity of the more massive planet. Again, the different line colors correspond to different mass ratios, and the left(right) panel correspond to a large(small) mass of the more massive planet. Here we can see that the locus of the neck shifts to higher eccentricities of the less massive planet as the more massive planet's eccentricity increases (cf. Fig. \ref{fig:deltaa}). For $m_2/m_1\leq 1$, the neck is shifted to eccentricities larger than the more massive planets's eccentricity, but this trend saturates at $e\sim0.3$. For $m_2/m_1 >1$, the neck is always shifted to eccentricities smaller than the more massive planet's eccentricity. This behavior is independent of the mass of the more massive planet, which would be expected if the neck is associated to a change of topology as in the case of an equilibrium bifurcation.

\section{Conclusions}\label{sec:conclusions}

In this work, we have addressed the dynamics of the planetary 2:1 MMR through the technique of dynamical maps. {These maps have been constructed from short term simulations of the dynamical evolution.} We have applied this technique to three real planetary systems, which are good representatives of resonant planets pairs covering a wide range of mass ratios. We have also applied the semi-analytical model of GBG21, that resulted to provide a very good agreement to the numerical simulations, and helped to understand the underlying dynamics observed in the maps. Our conclusions can be summarized as follows:
\begin{itemize}
\item The family of stable resonant equilibria may bifurcate from symmetric to asymmetric librations, depending on the planets' and eccentricities and mass ratios. The bifurcation, however, does not depend on the individual masses.
\item The semi-analytic width of the MMR fits very well, in general, to the outcomes of the dynamical maps
\item For mass ratios $m_2/m_1> 1$, increasing the eccentricity $e_2$ of the perturbing planet (i.e. the more massive planet) causes a bifurcation of the resonance center in $\sigma_1$, passing from one symmetric point at high eccentricities $e_1$ of the perturbed (less massive) planet, to two asymmetric points at low eccentricities.
\item For mass ratios $m_2/m_1< 1$, increasing the eccentricity $e_1$ of the perturbing planet causes also a bifurcation of the resonance center in $\sigma_2$, passing from one symmetric point at low eccentricities $e_2$ of the perturbed planet, to two asymmetric points at high eccentricities.
\item The occurrence of asymmetric librations must be taken into account in constructing the dynamical maps, in order to correctly assess the behavior of the system.
\item The resonant domain in the $a_i,e_i$ space, when considering non zero eccentricities $e_j$ of the more massive planet, is represented by a sand clock shape. The neck of the sand clock, i.e. the region where the width $\delta a_i$ is the smallest, is related to the bifurcations of the resonant centers of $\sigma_i$.
\item Within the uncertainties of the current orbital best fit, the TOI-216 system is locked in the 2:1 MMR, independently of the values of $\Delta\varpi$.
\item The HD27894 system would be locked in a 2:1 resonant libration state only if $\Delta\varpi=180^\circ$, which is not the solution supported by the observations.
\item Within the uncertainties of the current orbital fit, the K2-24 system is not locked in the 2:1 MMR, independently of the values of $\Delta\varpi$.
\end{itemize}

\section*{Acknowledgements}
The simulations presented in this work have been run at the SDumont and LoboC clusters of the Brazilian System of High-Performance Computing (SINAPAD), and at the Mulatona Cluster from the CCAD-UNC, which is part of SNCAD-MinCyT, Argentina. FR wishes to acknowledge support from the Brazilian National Council of Research (CNPq). This work has been partially financed by the Coordena\c{c}\~ao de Aperfei\c{c}oamento de Pessoal de N\'{\i}vel Superior – Brasil (CAPES) – Finance Code 001.  


\section*{Data availability}
The data underlying this article will be shared on reasonable request to the corresponding author.

\section*{Declarations}

\begin{itemize}
\item Funding: Not applicable.
\item Competing interests: There are no competing interest related to this work
\item Ethics approval: Not applicable
\item Consent to participate: Not applicable
\item Consent for publication: Not applicable
\item Availability of data and materials: Not applicable
\item Code availability: Not applicable
\item Authors' contributions: Not applicable
\end{itemize}

\bibliographystyle{mnras}
\bibliography{mmr21} 

\begin{thebibliography}{}
\makeatletter
\relax
\def\mn@urlcharsother{\let\do\@makeother \do\$\do\&\do\#\do\^\do\_\do\%\do\~}
\def\mn@doi{\begingroup\mn@urlcharsother \@ifnextchar [ {\mn@doi@}
  {\mn@doi@[]}}
\def\mn@doi@[#1]#2{\def\@tempa{#1}\ifx\@tempa\@empty \href
  {http://dx.doi.org/#2} {doi:#2}\else \href {http://dx.doi.org/#2} {#1}\fi
  \endgroup}
\def\mn@eprint#1#2{\mn@eprint@#1:#2::\@nil}
\def\mn@eprint@arXiv#1{\href {http://arxiv.org/abs/#1} {{\tt arXiv:#1}}}
\def\mn@eprint@dblp#1{\href {http://dblp.uni-trier.de/rec/bibtex/#1.xml}
  {dblp:#1}}
\def\mn@eprint@#1:#2:#3:#4\@nil{\def\@tempa {#1}\def\@tempb {#2}\def\@tempc
  {#3}\ifx \@tempc \@empty \let \@tempc \@tempb \let \@tempb \@tempa \fi \ifx
  \@tempb \@empty \def\@tempb {arXiv}\fi \@ifundefined
  {mn@eprint@\@tempb}{\@tempb:\@tempc}{\expandafter \expandafter \csname
  mn@eprint@\@tempb\endcsname \expandafter{\@tempc}}}

\bibitem[\protect\citeauthoryear{{Anglada-Escud{\'e}}, {L{\'o}pez-Morales}  \&
  {Chambers}}{{Anglada-Escud{\'e}} et~al.}{2010}]{Escude2010}
{Anglada-Escud{\'e}} G.,  {L{\'o}pez-Morales} M.,   {Chambers} J.~E.,  2010,
  \mn@doi [\apj] {10.1088/0004-637X/709/1/168}, \href
  {https://ui.adsabs.harvard.edu/abs/2010ApJ...709..168A} {709, 168}

\bibitem[\protect\citeauthoryear{{Batygin} \& {Adams}}{{Batygin} \&
  {Adams}}{2017}]{Batygin2017}
{Batygin} K.,  {Adams} F.~C.,  2017, \mn@doi [\aj]
  {10.3847/1538-3881/153/3/120}, \href
  {https://ui.adsabs.harvard.edu/abs/2017AJ....153..120B} {153, 120}

\bibitem[\protect\citeauthoryear{{Batygin} \& {Morbidelli}}{{Batygin} \&
  {Morbidelli}}{2013}]{Batygin2013b}
{Batygin} K.,  {Morbidelli} A.,  2013, \mn@doi [\aap]
  {10.1051/0004-6361/201220907}, \href
  {https://ui.adsabs.harvard.edu/abs/2013A&A...556A..28B} {556, A28}

\bibitem[\protect\citeauthoryear{{Bean} \& {Seifahrt}}{{Bean} \&
  {Seifahrt}}{2009}]{Bean2009}
{Bean} J.~L.,  {Seifahrt} A.,  2009, \mn@doi [\aap]
  {10.1051/0004-6361/200811280}, \href
  {https://ui.adsabs.harvard.edu/abs/2009A&A...496..249B} {496, 249}

\bibitem[\protect\citeauthoryear{{Beauge}}{{Beauge}}{1994}]{Beauge1994}
{Beauge} C.,  1994, \mn@doi [Celestial Mechanics and Dynamical Astronomy]
  {10.1007/BF00693323}, \href
  {https://ui.adsabs.harvard.edu/abs/1994CeMDA..60..225B} {60, 225}

\bibitem[\protect\citeauthoryear{{Beaug{\'e}}, {Ferraz-Mello}  \&
  {Michtchenko}}{{Beaug{\'e}} et~al.}{2003}]{Beauge2003}
{Beaug{\'e}} C.,  {Ferraz-Mello} S.,   {Michtchenko} T.~A.,  2003, \mn@doi
  [\apj] {10.1086/376568}, \href
  {https://ui.adsabs.harvard.edu/abs/2003ApJ...593.1124B} {593, 1124}

\bibitem[\protect\citeauthoryear{{Beaug{\'e}}, {Michtchenko}  \&
  {Ferraz-Mello}}{{Beaug{\'e}} et~al.}{2006}]{Beauge2006}
{Beaug{\'e}} C.,  {Michtchenko} T.~A.,   {Ferraz-Mello} S.,  2006, \mn@doi
  [\mnras] {10.1111/j.1365-2966.2005.09779.x}, \href
  {https://ui.adsabs.harvard.edu/abs/2006MNRAS.365.1160B} {365, 1160}

\bibitem[\protect\citeauthoryear{{Beaug{\'e}}, {Giuppone}, {Ferraz-Mello}  \&
  {Michtchenko}}{{Beaug{\'e}} et~al.}{2008}]{Beauge2008}
{Beaug{\'e}} C.,  {Giuppone} C.~A.,  {Ferraz-Mello} S.,   {Michtchenko} T.~A.,
  2008, \mn@doi [\mnras] {10.1111/j.1365-2966.2008.12979.x}, \href
  {https://ui.adsabs.harvard.edu/abs/2008MNRAS.385.2151B} {385, 2151}

\bibitem[\protect\citeauthoryear{{Boisvert}, {Nelson}  \& {Steffen}}{{Boisvert}
  et~al.}{2018}]{Boisvert+2018}
{Boisvert} J.~H.,  {Nelson} B.~E.,   {Steffen} J.~H.,  2018, \mn@doi [\mnras]
  {10.1093/mnras/sty2023}, \href
  {https://ui.adsabs.harvard.edu/abs/2018MNRAS.480.2846B} {480, 2846}

\bibitem[\protect\citeauthoryear{{Charalambous}, {Ramos},
  {Ben{\'\i}tez-Llambay}  \& {Beaug{\'e}}}{{Charalambous}
  et~al.}{2017}]{Charalambous+2018}
{Charalambous} C.,  {Ramos} X.~S.,  {Ben{\'\i}tez-Llambay} P.,   {Beaug{\'e}}
  C.,  2017, in Journal of Physics Conference Series. p. 012027

\bibitem[\protect\citeauthoryear{{Chatterjee} \& {Ford}}{{Chatterjee} \&
  {Ford}}{2015}]{Chatterjee2015}
{Chatterjee} S.,  {Ford} E.~B.,  2015, \mn@doi [\apj]
  {10.1088/0004-637X/803/1/33}, \href
  {https://ui.adsabs.harvard.edu/abs/2015ApJ...803...33C} {803, 33}

\bibitem[\protect\citeauthoryear{{Cincotta} \& {Sim{\'o}}}{{Cincotta} \&
  {Sim{\'o}}}{2000}]{Cincotta.Simo.1999}
{Cincotta} P.~M.,  {Sim{\'o}} C.,  2000, \mn@doi [\aaps] {10.1051/aas:2000108},
  \href {http://adsabs.harvard.edu/abs/2000A%26AS..147..205C} {147, 205}

\bibitem[\protect\citeauthoryear{{Cincotta}, {Giordano}  \&
  {Sim{\'o}}}{{Cincotta} et~al.}{2003}]{Cincotta2003}
{Cincotta} P.~M.,  {Giordano} C.~M.,   {Sim{\'o}} C.,  2003, \mn@doi [Physica D
  Nonlinear Phenomena] {10.1016/S0167-27890300103-9}, \href
  {https://ui.adsabs.harvard.edu/abs/2003PhyD..182..151C} {182, 151}

\bibitem[\protect\citeauthoryear{{Dai} et~al.,}{{Dai} et~al.}{2016}]{Dai+2016}
{Dai} F.,  et~al., 2016, \mn@doi [\apj] {10.3847/0004-637X/823/2/115}, \href
  {https://ui.adsabs.harvard.edu/abs/2016ApJ...823..115D} {823, 115}

\bibitem[\protect\citeauthoryear{{Dawson} et~al.,}{{Dawson}
  et~al.}{2019}]{dawson19}
{Dawson} R.~I.,  et~al., 2019, \mn@doi [\aj] {10.3847/1538-3881/ab24ba}, \href
  {https://ui.adsabs.harvard.edu/abs/2019AJ....158...65D} {158, 65}

\bibitem[\protect\citeauthoryear{{Dawson} et~al.,}{{Dawson}
  et~al.}{2021}]{dawson21}
{Dawson} R.~I.,  et~al., 2021, \mn@doi [\aj] {10.3847/1538-3881/abd8d0}, \href
  {https://ui.adsabs.harvard.edu/abs/2021AJ....161..161D} {161, 161}

\bibitem[\protect\citeauthoryear{{Deck}, {Payne}  \& {Holman}}{{Deck}
  et~al.}{2013}]{Deck2013}
{Deck} K.~M.,  {Payne} M.,   {Holman} M.~J.,  2013, \mn@doi [\apj]
  {10.1088/0004-637X/774/2/129}, \href
  {https://ui.adsabs.harvard.edu/abs/2013ApJ...774..129D} {774, 129}

\bibitem[\protect\citeauthoryear{{Delisle} \& {Laskar}}{{Delisle} \&
  {Laskar}}{2014}]{Delisle2014}
{Delisle} J.~B.,  {Laskar} J.,  2014, \mn@doi [\aap]
  {10.1051/0004-6361/201424227}, \href
  {https://ui.adsabs.harvard.edu/abs/2014A&A...570L...7D} {570, L7}

\bibitem[\protect\citeauthoryear{{Delisle}, {Laskar}  \& {Correia}}{{Delisle}
  et~al.}{2014}]{Delisle2014A}
{Delisle} J.~B.,  {Laskar} J.,   {Correia} A.~C.~M.,  2014, \mn@doi [\aap]
  {10.1051/0004-6361/201423676}, \href
  {https://ui.adsabs.harvard.edu/abs/2014A&A...566A.137D} {566, A137}

\bibitem[\protect\citeauthoryear{{Gallardo}, {Beaug{\'e}}  \&
  {Giuppone}}{{Gallardo} et~al.}{2021}]{Gallardo2021}
{Gallardo} T.,  {Beaug{\'e}} C.,   {Giuppone} C.~A.,  2021, \mn@doi [\aap]
  {10.1051/0004-6361/202039764}, \href
  {https://ui.adsabs.harvard.edu/abs/2021A&A...646A.148G} {646, A148}

\bibitem[\protect\citeauthoryear{{Giuppone}, {Tadeu dos Santos}, {Beaug{\'e}},
  {Ferraz-Mello}  \& {Michtchenko}}{{Giuppone} et~al.}{2009}]{Giuppone2009}
{Giuppone} C.~A.,  {Tadeu dos Santos} M.,  {Beaug{\'e}} C.,  {Ferraz-Mello} S.,
    {Michtchenko} T.~A.,  2009, \mn@doi [\apj] {10.1088/0004-637X/699/2/1321},
  \href {https://ui.adsabs.harvard.edu/abs/2009ApJ...699.1321G} {699, 1321}

\bibitem[\protect\citeauthoryear{{Giuppone}, {Ben{\'\i}tez-Llambay}  \&
  {Beaug{\'e}}}{{Giuppone} et~al.}{2012}]{Giuppone2012}
{Giuppone} C.~A.,  {Ben{\'\i}tez-Llambay} P.,   {Beaug{\'e}} C.,  2012, \mn@doi
  [\mnras] {10.1111/j.1365-2966.2011.20310.x}, \href
  {https://ui.adsabs.harvard.edu/abs/2012MNRAS.421..356G} {421, 356}

\bibitem[\protect\citeauthoryear{{Goldreich} \& {Schlichting}}{{Goldreich} \&
  {Schlichting}}{2014}]{Goldreich+2014}
{Goldreich} P.,  {Schlichting} H.~E.,  2014, \mn@doi [\aj]
  {10.1088/0004-6256/147/2/32}, \href
  {https://ui.adsabs.harvard.edu/abs/2014AJ....147...32G} {147, 32}

\bibitem[\protect\citeauthoryear{{Hinse}, {Christou}, {Alvarellos}  \&
  {Go{\'z}dziewski}}{{Hinse} et~al.}{2010}]{Hinse2010}
{Hinse} T.~C.,  {Christou} A.~A.,  {Alvarellos} J.~L.~A.,   {Go{\'z}dziewski}
  K.,  2010, \mn@doi [\mnras] {10.1111/j.1365-2966.2010.16307.x}, \href
  {https://ui.adsabs.harvard.edu/abs/2010MNRAS.404..837H} {404, 837}

\bibitem[\protect\citeauthoryear{{Kipping}, {Nesvorn{\'y}}, {Hartman},
  {Torres}, {Bakos}, {Jansen}  \& {Teachey}}{{Kipping}
  et~al.}{2019}]{kipping19}
{Kipping} D.,  {Nesvorn{\'y}} D.,  {Hartman} J.,  {Torres} G.,  {Bakos} G.,
  {Jansen} T.,   {Teachey} A.,  2019, MNRAS, 486, 4980

\bibitem[\protect\citeauthoryear{{K{\"u}rster}, {Trifonov}, {Reffert},
  {Kostogryz}  \& {Rodler}}{{K{\"u}rster} et~al.}{2015}]{Kurster2015}
{K{\"u}rster} M.,  {Trifonov} T.,  {Reffert} S.,  {Kostogryz} N.~M.,   {Rodler}
  F.,  2015, \mn@doi [\aap] {10.1051/0004-6361/201525872}, \href
  {https://ui.adsabs.harvard.edu/abs/2015A&A...577A.103K} {577, A103}

\bibitem[\protect\citeauthoryear{{Michtchenko}, {Beaug{\'e}}  \&
  {Ferraz-Mello}}{{Michtchenko} et~al.}{2008a}]{Michtchenko2008a}
{Michtchenko} T.~A.,  {Beaug{\'e}} C.,   {Ferraz-Mello} S.,  2008a, \mn@doi
  [\mnras] {10.1111/j.1365-2966.2008.13278.x}, \href
  {https://ui.adsabs.harvard.edu/abs/2008MNRAS.387..747M} {387, 747}

\bibitem[\protect\citeauthoryear{{Michtchenko}, {Beaug{\'e}}  \&
  {Ferraz-Mello}}{{Michtchenko} et~al.}{2008b}]{Michtchenko2008b}
{Michtchenko} T.~A.,  {Beaug{\'e}} C.,   {Ferraz-Mello} S.,  2008b, \mn@doi
  [\mnras] {10.1111/j.1365-2966.2008.13867.x}, \href
  {https://ui.adsabs.harvard.edu/abs/2008MNRAS.391..215M} {391, 215}

\bibitem[\protect\citeauthoryear{{Moutou} et~al.,}{{Moutou}
  et~al.}{2005}]{Moutou2005}
{Moutou} C.,  et~al., 2005, \mn@doi [\aap] {10.1051/0004-6361:20052826}, \href
  {https://ui.adsabs.harvard.edu/abs/2005A&A...439..367M} {439, 367}

\bibitem[\protect\citeauthoryear{{Nesvorn{\'y}}, {Chrenko}  \&
  {Flock}}{{Nesvorn{\'y}} et~al.}{2022}]{Nesvorny2022}
{Nesvorn{\'y}} D.,  {Chrenko} O.,   {Flock} M.,  2022, \mn@doi [\apj]
  {10.3847/1538-4357/ac36cd}, \href
  {https://ui.adsabs.harvard.edu/abs/2022ApJ...925...38N} {925, 38}

\bibitem[\protect\citeauthoryear{{Petigura} et~al.,}{{Petigura}
  et~al.}{2016}]{Petigura+2016}
{Petigura} E.~A.,  et~al., 2016, \mn@doi [\apj] {10.3847/0004-637X/818/1/36},
  \href {https://ui.adsabs.harvard.edu/abs/2016ApJ...818...36P} {818, 36}

\bibitem[\protect\citeauthoryear{{Petigura} et~al.,}{{Petigura}
  et~al.}{2018}]{Petigura+2018}
{Petigura} E.~A.,  et~al., 2018, \mn@doi [\aj] {10.3847/1538-3881/aaceac},
  \href {https://ui.adsabs.harvard.edu/abs/2018AJ....156...89P} {156, 89}

\bibitem[\protect\citeauthoryear{{Press}, {Teukolsky}, {Vetterling}  \&
  {Flannery}}{{Press} et~al.}{1992}]{NumRecip1992}
{Press} W.~H.,  {Teukolsky} S.~A.,  {Vetterling} W.~T.,   {Flannery} B.~P.,
  1992, {Numerical recipes in FORTRAN. The art of scientific computing}.
{Cambridge University Press}

\bibitem[\protect\citeauthoryear{{Ramos}, {Charalambous},
  {Ben{\'\i}tez-Llambay}  \& {Beaug{\'e}}}{{Ramos} et~al.}{2017}]{Ramos+2017}
{Ramos} X.~S.,  {Charalambous} C.,  {Ben{\'\i}tez-Llambay} P.,   {Beaug{\'e}}
  C.,  2017, \mn@doi [\aap] {10.1051/0004-6361/201629642}, \href
  {https://ui.adsabs.harvard.edu/abs/2017A&A...602A.101R} {602, A101}

\bibitem[\protect\citeauthoryear{{Rivera}, {Laughlin}, {Butler}, {Vogt},
  {Haghighipour}  \& {Meschiari}}{{Rivera} et~al.}{2010}]{Rivera2010}
{Rivera} E.~J.,  {Laughlin} G.,  {Butler} R.~P.,  {Vogt} S.~S.,  {Haghighipour}
  N.,   {Meschiari} S.,  2010, \mn@doi [\apj] {10.1088/0004-637X/719/1/890},
  \href {https://ui.adsabs.harvard.edu/abs/2010ApJ...719..890R} {719, 890}

\bibitem[\protect\citeauthoryear{{Trifonov}, {Reffert}, {Tan}, {Lee}  \&
  {Quirrenbach}}{{Trifonov} et~al.}{2014}]{Trifonov2014}
{Trifonov} T.,  {Reffert} S.,  {Tan} X.,  {Lee} M.~H.,   {Quirrenbach} A.,
  2014, \mn@doi [\aap] {10.1051/0004-6361/201322885}, \href
  {https://ui.adsabs.harvard.edu/abs/2014A&A...568A..64T} {568, A64}

\bibitem[\protect\citeauthoryear{{Trifonov} et~al.,}{{Trifonov}
  et~al.}{2017}]{Trifonov+2017}
{Trifonov} T.,  et~al., 2017, \mn@doi [\aap] {10.1051/0004-6361/201731044},
  \href {https://ui.adsabs.harvard.edu/abs/2017A&A...602L...8T} {602, L8}

\bibitem[\protect\citeauthoryear{{Voyatzis}, {Kotoulas}  \&
  {Hadjidemetriou}}{{Voyatzis} et~al.}{2009}]{Voyatzis2009}
{Voyatzis} G.,  {Kotoulas} T.,   {Hadjidemetriou} J.~D.,  2009, \mn@doi
  [\mnras] {10.1111/j.1365-2966.2009.14671.x}, \href
  {https://ui.adsabs.harvard.edu/abs/2009MNRAS.395.2147V} {395, 2147}

\makeatother
\end{thebibliography}


\section{Appendix}\label{Append-1}

\textbf{Additional maps for TOI-216 and HD27894}
\begin{figure*}
    \centering
    \includegraphics[width=0.49\textwidth]{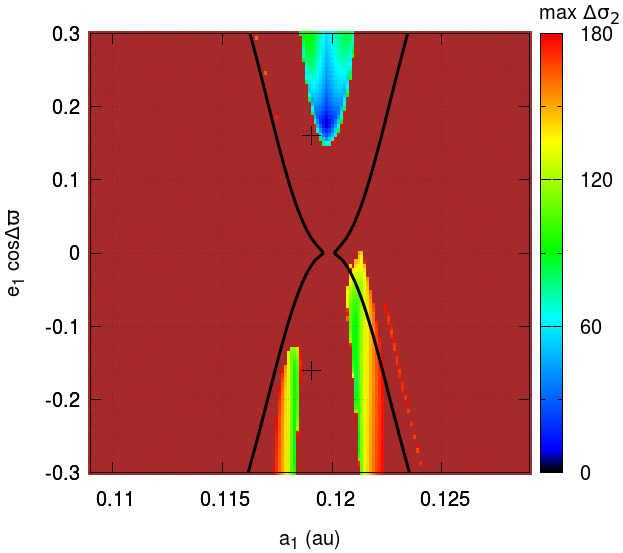}
    \includegraphics[width=0.49\textwidth]{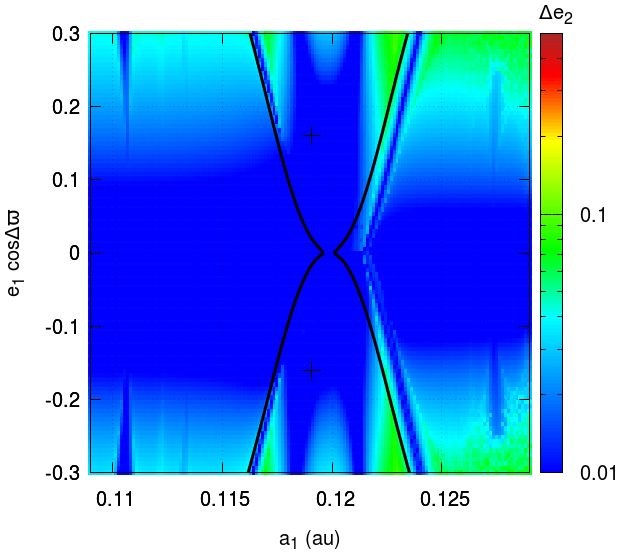}
    \caption{Dynamical maps for the TOI-216 system in the plane $a_1, e_1 cos(\Delta \varpi$), showing the variations of the more massive planet, $\max\Delta\sigma_2$ and $\max\Delta e_2$. Panels shall be compared to Fig. \ref{fig:toi-216}. {It is noteworthy the libration of $\sigma_2$ for values of $e_1\gtrsim 0.15$, which, combined with the libration of $\sigma_1$, is consistent with the presence of ACRs in that region \citep[e.g.][]{Michtchenko2008a}}}
    \label{fig:toi216-massive}
\end{figure*}
\begin{figure*}
    \centering
    \includegraphics[width=0.49\textwidth]{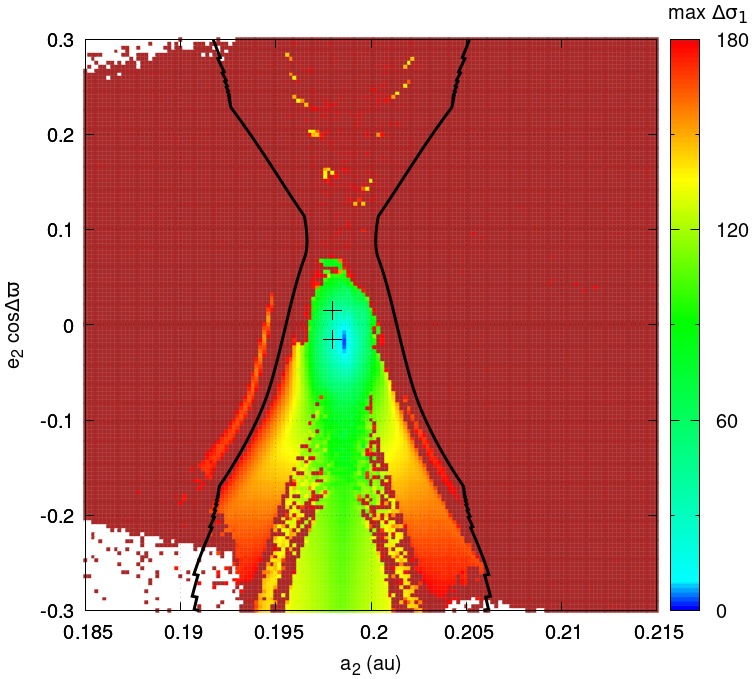}
    \includegraphics[width=0.49\textwidth]{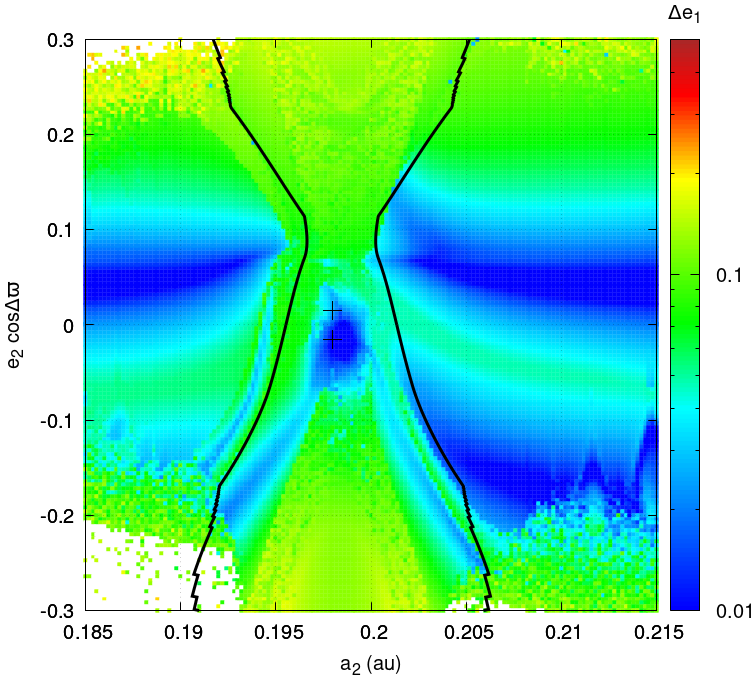}
    \caption{Dynamical maps for the HD27894 system in the plane $a_2, e_2 cos(\Delta \varpi$), showing the variations of the more massive planet, $\max\Delta\sigma_1$ and $\max\Delta e_1$. Panels shall be compared to Fig. \ref{fig:hd27894}. {The libration of $\sigma_1$ observed here for the whole range of $e_2\Delta\varpi<0$, combined with the libration of $\sigma_2$ about $e_2\Delta\varpi\sim -0.02$, implies in the occurrence of ACRs consistent which the theoretical predictions for the mass ratio considered \citep[e.g.][]{Michtchenko2008b}.}}
    \label{fig:hd27894-massive}
\end{figure*}

\bsp	
\label{lastpage}
\end{document}